\documentclass[]{pasj02} 
\usepackage[switch,mathlines]{lineno}

\jyear{2026}
\Received{}
\Accepted{}

\begin{document}


\title{The Milky Way Tomography with Subaru Hyper Suprime-Cam. II. Global halo structure}

\author{
Yoshihisa \textsc{Suzuki}\altaffilmark{1}, \email{yoshihisa.suzuki@astr.tohoku.ac.jp} 
Masashi \textsc{Chiba}\altaffilmark{1},
Rosemary F. G. \textsc{Wyse}\altaffilmark{2},
Shunichi \textsc{Horigome}\altaffilmark{1}
}
\altaffiltext{1}{Astronomical Institute, Tohoku University, Aoba-ku, Sendai 980-8578, Japan}
\altaffiltext{2}{The William H. Miller III, Dept. of Physics \& Astronomy, Johns Hopkins University, Baltimore, MD 21218, USA}


\maketitle


\begin{abstract}
We investigate the structure of the Milky Way's stellar halo within 70~kpc of the Sun using a wide-field photometric catalog obtained from the Hyper Suprime-Cam (HSC) Subaru Strategic Program (HSC-SSP).
We employ a large sample of main-sequence turn-off stars as distance tracers.
To robustly derive the structural parameters of the stellar halo, we develop a forward-modeling framework that explicitly accounts for distance uncertainties, the solar position, and the limited sky coverage of the survey.
Applying this method to the HSC-SSP catalog, we found that the smooth stellar halo is well described by a double power-law density profile, with inner and outer slope of approximately -3.3 and -4.8, respectively, with a break radius of 17.4~kpc. 
The outer steep density slope derived in this work supports a picture in which the present-day structure of the Milky Way's stellar halo is influenced by early massive accretion events, consistent with inferences from kinematic substructures such as Gaia Enceladus/Sausage.
Ongoing wide-field imaging surveys, including UNIONS and LSST, will provide further constraints on the structure of the stellar halo and key insights into its formation history.
\end{abstract}


\section{Introduction}
In the $\Lambda$CDM paradigm, galaxies such as the Milky Way (MW) are assembled through successive mergers and accretions of dark matter subhalos and associated baryons \citep{White+1978}. 
Cosmological simulations predict that the stellar halo, predominantly composed of stars deposited from disrupted smaller systems such as dwarf galaxies and globular clusters, preserves long-lived signatures of these assembly processes. 
Such signatures are imprinted both in the global density profile of the halo and in a variety of substructures, including tidal streams, shells, and overdensities \citep{Bullock+2005, Springel+2005, Cooper+2010}.
Halo structures therefore offer a valuable fossil record of the assembly history of a galaxy.

The MW's stellar halo has long been studied as a key probe of galaxy formation. 
\citet{Eggen+1962} advocated a rapid, monolithic collapse of the proto-Galaxy, whereas \citet{Searle+1978} proposed successive accretion of smaller stellar systems, based on the phase-space and chemical abundance information of old stars.
In the past two decades, subsequent wide-field photometric surveys such as the Sloan Digital Sky Survey (SDSS) \citep{York+2000}, Panoramic Survey Telescope and Rapid Response System (Pan-STARRS) \citep{Chambers+2016}, Dark Energy Survey (DES) \citep{Abbott+2018} have revealed that the stellar halo is highly structured, with numerous overdensities identified through star-count analyses.
Following spectroscopic programs, including the Sloan Extension for Galactic Understanding and Exploration (SEGUE) \citep{Yanny+2009} and LAMOST \citep{Zhao+2012}, further revealed the presence of a dual halo structure from a chemo-dynamical perspective \citep{Carollo+2007}.
Recently, Gaia has provided six-dimensional phase-space information for millions of stars, enabling the identification of major accretion events such as Gaia--Enceladus/Sausage \citep{Helmi+2018, Belokurov+2018} and additional events including Sequoia \citep{Koppelman+2019} as the kinematic substructures.
These studies have established the MW as a unique laboratory in which the assembly history of a stellar halo can be examined in exceptional detail.

Mapping the stellar halo over a wide range of Galactocentric distances requires reliable stellar distance tracers.
Various stellar populations have been employed for this purpose, each characterized by a trade-off between the precision and accuracy of distance measurements and number density of tracers.
RR Lyrae stars and blue horizontal branch (BHB) stars provide relatively precise distance estimates, with typical uncertainties of $\sim$10--15\%, owing to their well-calibrated luminosities \citep{Watkins+2009, Deason+2011, Fukushima+2025}.
Another commonly used tracer population is K-giants, which occupy the red giant branch phase of stellar evolution and exhibit significant luminosity dispersion at fixed color or temperature, resulting in distance uncertainties that are generally larger than those of RR Lyrae or BHB stars \citep{Xue+2015}.
However, these latter bright tracers are intrinsically rare, which limits their ability to trace low-surface-brightness halo structures with wide angular extents.
That said, several outer-halo substructures have been identified using such sparse tracers.
For example, the Pisces Overdensity was detected at a heliocentric distance of $\sim$80~kpc using RR Lyrae stars \citep{Sesar+2007, Watkins+2009}, and has also been studied with BHB stars \citep{Nie+2015}.
The region toward Virgo is particularly complex, hosting multiple overdensities.
The Virgo Overdensity lies in the inner halo at $\sim$10--20~kpc \citep{Vivas+2001, Juric+2008, Carlin+2012}, while a more distant structure, often referred to as the Outer Virgo Overdensity, has been reported at $\sim$80--120~kpc based on RR Lyrae stars \citep{Sesar+2017}.
Despite these successes, the limited number statistics of such tracers hinder a homogeneous and statistically robust characterization of the stellar halo, especially in its outer regions.

In this context, main-sequence turn-off (MSTO) stars offer a complementary approach to tracing the stellar halo.
Because MSTO stars are far more numerous than classical halo standard candles, they enable the detection of diffuse and extended halo structures that are otherwise difficult to identify.
Indeed, large-scale mapping of MSTO stars has revealed the global shape of the stellar halo within 20 kpc and uncovered numerous substructures, including tidal streams and overdensities \citep{Belokurov+2006, Juric+2008}.
However, the absolute magnitudes of MSTO stars depend sensitively on stellar age and metallicity.
As a result, distance estimates for individual MSTO stars only from broad-band photometry are subject to substantial uncertainties, typically at the level of $\sim$20--30\% \citep{Newberg+2006, Bell+2008}. 
Consequently, while MSTO stars are statistically powerful tracers of the stellar halo, extracting quantitative information on the smooth halo structure requires careful distance estimation and a modeling framework that explicitly accounts for these uncertainties.

This paper is the second in the series \textit{The Milky Way Tomography with Subaru Hyper Suprime-Cam}.
The primary objective of this work is to establish a robust methodology for tracing the stellar halo using MSTO stars by combining a well-defined distance estimation method with a statistical framework that incorporates distance uncertainties into the inference of halo structure.
Using deep and wide-field photometry from the Hyper Suprime-Cam Subaru Strategic Program (HSC-SSP), we derive distances to MSTO stars and infer the structural parameters of the smooth stellar halo while properly accounting for distance errors.

This paper is organized as follows.
Section~2 briefly describes the HSC-SSP imaging data and the selection of halo stars, then describes the MSTO distance estimation method and the modeling framework used to infer the halo structure, including the treatment of distance uncertainties.
Section~3 presents the map of MSTO stars and the resulting halo parameters.
Section~4 discusses their implications for the assembly history of the MW.
Our conclusions are summarized in Section~5.


\begin{figure*}[h!]
    \begin{center}
    \includegraphics[width=\textwidth, keepaspectratio]{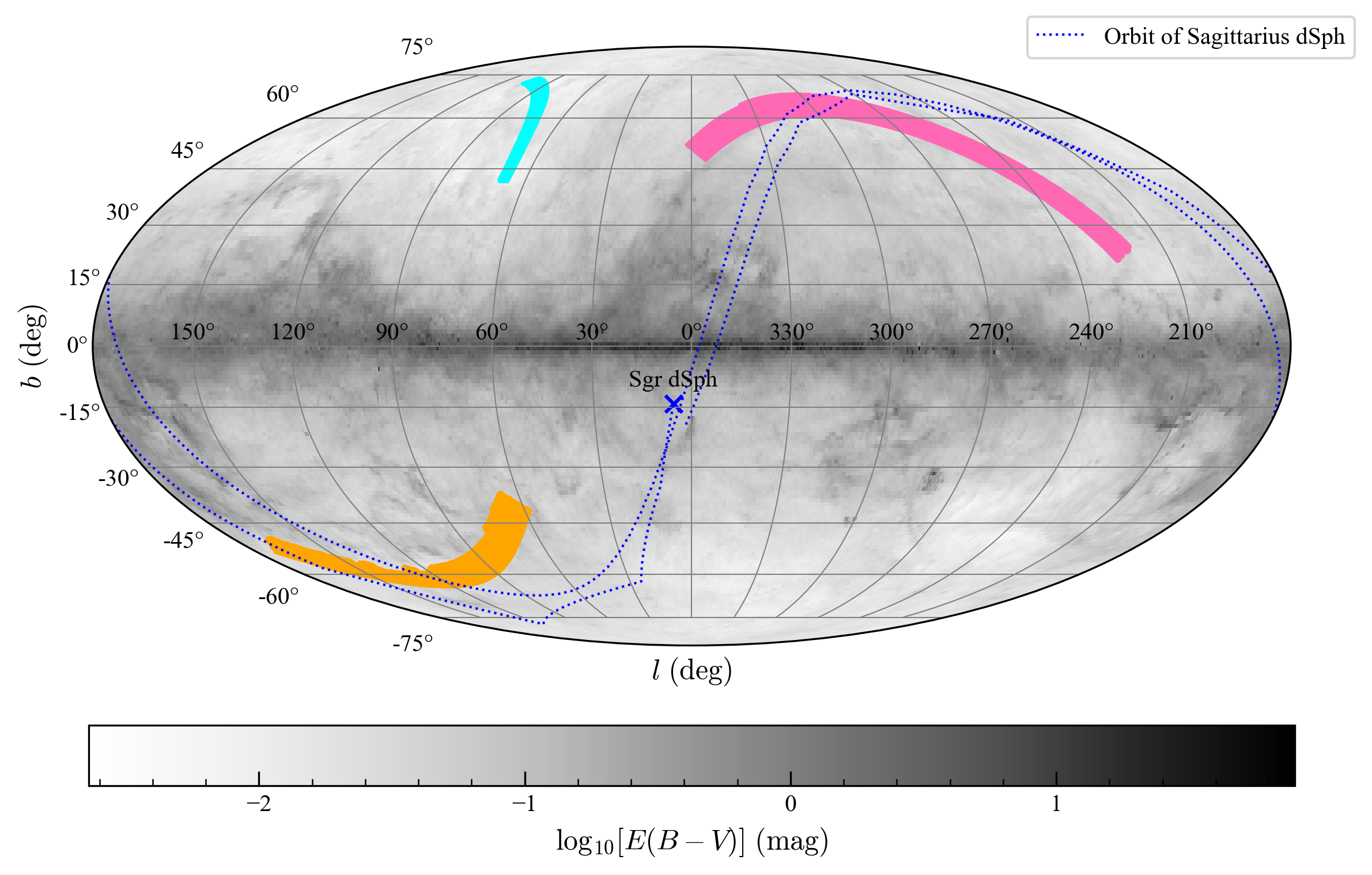}
    \end{center}
    \caption{
    Survey footprint of the Wide layer of the HSC-SSP in Galactic coordinates.
    The Spring, Fall, and North fields are shown in pink, orange, and cyan, respectively.
    The blue dotted curve indicates the model of past orbit of the Sagittarius dwarf spheroidal galaxy based on \citet{Vasiliev+2021}.
    The background shows the extinction map from \citet{Schlegel+1998}.
    {Alt text: A sky map in Galactic coordinates showing the survey footprint of the HSC-SSP Wide layer. The layer consists of three separated fields labeled Spring, Fall, and North, located at different Galactic longitudes and at relatively high Galactic latitudes, away from the Galactic plane.}
    }
    \label{fig1}
\end{figure*}

\section{Data and Method}
In this section, we describe the imaging dataset and the overall methodology adopted in this work.
We first briefly introduce the HSC-SSP and summarize the construction of the halo star sample.
All details of the star--galaxy separation and halo star selection are presented in Paper~I of this series, and we adopt the same sample definition throughout this work.
Finally, we present the framework used to infer the structural parameters of the stellar halo in a Galactocentric coordinate system, explicitly accounting for distance uncertainties, the solar position, and the limited sky coverage of the survey.

\subsection{HSC-SSP}
We use the imaging catalog from the Wide layer (S21A internal data release) of the HSC-SSP.
The catalog was generated using \texttt{hscPipe} version 8.4, with photometric calibration updated by the Forward Global Calibration Method (FGCM; \texttt{hscPipe} 8.5.3).
The pipeline is based on that developed for the Vera C. Rubin Observatory \citep{Ivezic+2008, Juric+2017, Bosch+2019}, and the photometry and astrometry are calibrated against Pan-STARRS1 \citep{Schlafly+2012, Tonry+2012, Magnier+2013, Chambers+2016}.

The HSC-SSP Wide layer provides deep multi-band photometry in five broad bands ($g$, $r$, $i$, $z$, and $y$) over a total area of approximately 1,110~deg$^{2}$.
Figure~\ref{fig1} shows the survey footprint of the Wide layer in Galactic coordinates.
The footprint consists of three spatially separated regions---the Spring, Fall, and North fields---whose geometries approximately follow lines of constant declination in the equatorial coordinate system.
These fields predominantly probe high Galactic latitudes, making them well suited for studies of the stellar halo while reducing contamination from disk populations.
The blue dotted curve indicates the projected orbit of the Sagittarius dwarf spheroidal galaxy \citep{Vasiliev+2021}, highlighting that the survey footprint intersects regions known to host prominent halo substructures.

\subsection{Selection of halo stars}
The halo star sample used in this work is identical to that defined in Paper~I.
In particular, all criteria for star--galaxy separation and color-based selection of halo stars are adopted without modification.
We therefore do not repeat the selection procedure here, and refer the reader to Paper~I for a detailed description and validation of the sample.

\subsection{MSTO catalog construction}
MSTO stars are defined as those at the point where stellar evolution causes them to leave the main sequence and evolve toward the subgiant branch.
MSTO stars are commonly used as distance tracers since they are abundant and occupy a relatively well-defined region in color–magnitude space for old stellar populations.
Previous studies estimated distances to MSTO stars using empirical, color-based calibrations \citep{Juric+2008, Bell+2008, PilaDiez+2015}. 
In these approaches, absolute magnitudes are assigned through fitted relations based on multi broad-band colors, and distances are obtained by comparing them with the observed apparent magnitudes.
Such methods do not explicitly encode the underlying stellar physics and do not account for the distribution of stellar masses along the main sequence.

We have developed an isochrone-based framework to infer photometric distances for MSTO stars.
Our method is conceptually related to isochrone-projection approaches \citep{Zwitter+2010, Kordopatis+2023}, but we make the dependence on the initial stellar mass along the isochrone explicit.
Specifically, we treat the initial mass as a latent variable and marginalize it using a physically motivated initial mass function (IMF).
By explicitly incorporating the IMF into the likelihood, our formulation provides a physically consistent description of the MSTO region, where stellar properties change rapidly with mass, and yields robust photometric distance estimates for mapping the three-dimensional structure of the Galactic halo.

\subsubsection{Definition}
We begin by defining the model parameters $\Theta$ and the observed data $\mathcal{D}$.
The set of physical parameters to be inferred is denoted by
\begin{equation}
\Theta \equiv (\tau, [\mathrm{M/H}], \mu),
\end{equation}
where $\tau$ is the stellar age, $[\mathrm{M/H}]$ is the metallicity, and $\mu$ is the distance modulus.
The observed data for each star are given by
\begin{equation}
\mathcal{D} \equiv \{ g_{\rm obs}, i_{\rm obs}, \sigma_g, \sigma_i, A_g, A_i \},
\end{equation}
where $g_{\rm obs}$ and $i_{\rm obs}$ are the observed apparent magnitudes in the $g$ and $i$ bands,
$A_g$ and $A_i$ are the extinction values in the corresponding bands assuming their uncertainties are neglected, and $\sigma_g$ and $\sigma_i$ are the photometric uncertainties.
The observed magnitudes in the $g$ and $i$ bands are modeled as
\begin{equation}
g_{\rm obs}=g_{\rm true}+\epsilon_g,\qquad
i_{\rm obs}=i_{\rm true}+\epsilon_i,
\end{equation}
with independent Gaussian errors
\begin{equation}
\epsilon_g\sim\mathcal{N}(0,\sigma_g^2), \qquad
\epsilon_i\sim\mathcal{N}(0,\sigma_i^2).
\end{equation}
We define extinction-corrected magnitudes as
\begin{equation}
g_0 = g_{\rm obs} - A_g, \qquad
i_0 = i_{\rm obs} - A_i .
\end{equation}

For a given set of parameters $\Theta$, the location of a star in the color--magnitude diagram depends not only on $(\tau, [\mathrm{M/H}], \mu)$ but also on its initial stellar mass $M_{\rm ini}$ along the isochrone.
We therefore define the likelihood for $\Theta$ by marginalizing over $M_{\rm ini}$,
\begin{equation}
L(\Theta \mid \mathcal{D})
=
\int p(\mathcal{D} \mid M_{\rm ini}, \Theta)\,\phi(M_{\rm ini})\,dM_{\rm ini},
\end{equation}
where $p(\mathcal{D} \mid M_{\rm ini}, \Theta)$ is the photometric likelihood in color--magnitude space and $\phi(M_{\rm ini})$ is the assumed IMF. 
Here, $\phi(M_{\rm ini})$ is used as a relative weighting along the isochrone, thus its absolute normalization does not affect the inferred parameters. 
Note that in such projection approaches \citep{Zwitter+2010, Kordopatis+2023}, the mass dependence along the isochrone is not explicitly weighted, which is equivalent to adopting a uniform prior in $M_{\rm ini}$.
By explicitly incorporating the IMF into the marginalization, our formulation yields a physically motivated likelihood.

\subsubsection{Distance inference framework}
We work in the color-magnitude space,
\begin{equation}
C_{\rm obs}=g_{0}-i_{0},\qquad
I_{\rm obs}=i_{0},
\end{equation}
where the corresponding uncertainties are
\begin{equation}
\sigma_C = \sqrt{\sigma_g^2+\sigma_i^2},\qquad
\sigma_I = \sigma_i,
\end{equation}
and the error covariance is
\begin{equation}
\Sigma =
\begin{pmatrix}
\sigma_g^2+\sigma_i^2 & -\sigma_i^2\\
-\sigma_i^2           & \sigma_i^2
\end{pmatrix}.
\end{equation}
We denote the components of the inverse covariance as
\begin{equation}
\Sigma^{-1}_{CC}=\sigma_g^{-2},\quad
\Sigma^{-1}_{CI}=\sigma_g^{-2},\quad
\Sigma^{-1}_{II}=\sigma_g^{-2}+\sigma_i^{-2}.
\end{equation}

For each age $\tau_a$ and metallicity $[\mathrm{M/H}]_m$, the isochrone provides absolute magnitudes 
$(g_k^{\rm iso}, i_k^{\rm iso})$ at discrete initial masses $M_{{\rm ini},k}$.  
A distance grid $D_d$ defines the distance modulus
\begin{equation}
\mu_d = 5\log_{10}(D_d/{\rm kpc}) + 10,
\end{equation}
and the predicted color and magnitude at grid point $(k,d)$ are
\begin{equation}
C_{k,d}=g_k^{\rm iso} - i_k^{\rm iso},\qquad
I_{k,d}=i_k^{\rm iso}+\mu_d.
\end{equation}
The residuals between the model and observed quantities are
\begin{equation}
\Delta C_{k,d}=C_{\rm obs}-C_{k,d},\qquad
\Delta I_{k,d}=I_{\rm obs}-I_{k,d},
\end{equation}

Based on these, we evaluate the likelihood only within a local region in color--magnitude space for computational efficiency.
The size of this region is controlled by a dimensionless parameter $n_\sigma$, which specifies how many times the 1$\sigma$ observational uncertainties are used to define the window.
In the following analysis, we choose $n_\sigma=$~5 to sufficiently capture the tails of the underlying distribution, such that contributions from the excluded region are statistically negligible.
We therefore evaluate the likelihood only for grid points satisfying
\begin{equation}
|\Delta C_{k,d}| \le n_\sigma\,\sigma_C,\qquad
|\Delta I_{k,d}| \le n_\sigma\,\sigma_I,
\end{equation}
and assign zero likelihood to grid points outside this window.
For accepted points, the local likelihood is
\begin{equation}
L_{k,d}\propto\exp\!\left(-\frac{\chi^2_{k,d}}{2}\right).
\end{equation}
where chi-square $\chi^2_{k,d}$ is
\begin{equation}
\chi^2_{k,d}
=
\Sigma^{-1}_{CC}\Delta C_{k,d}^2
+2\Sigma^{-1}_{CI}\Delta C_{k,d}\Delta I_{k,d}
+\Sigma^{-1}_{II}\Delta I_{k,d}^2.
\end{equation}

Then, marginalization over the initial mass is approximated by the discrete sum
\begin{equation}
L_{a,m,d}
\propto
\sum_k
\phi(M_{{\rm ini},k})\,
\Delta M_{{\rm ini},k}\,
L_{k,d},
\end{equation}
where $\Delta M_{{\rm ini},k}=M_{{\rm ini},k+1}-M_{{\rm ini},k}$ is the spacing of the isochrone in initial mass.

A prior $p(\tau_a, [\mathrm{M/H}]_m)$ is assigned to age and metallicity.
We adopt a uniform prior for distance modulus over the considered grid range.
The unnormalized posterior weight at grid point $(a,m,d)$ is therefore
\begin{equation}
w_{a,m,d}\propto L_{a,m,d}\,p(\tau_a,[\mathrm{M/H}]_m).
\end{equation}

For convenience, the grid indices are flattened into a single index $i$, storing the 
quantities $(\tau_i, [\mathrm{M/H}]_i, \mu_i, w_i)$.  
The posterior mean and variance of any parameter 
$\theta\in\{\tau, [\mathrm{M/H}], \mu\}$ are computed as
\begin{equation}
\langle \theta\rangle=\frac{\sum_i w_i \theta_i}{\sum_i w_i},\qquad
\sigma_\theta^2=\frac{\sum_i w_i(\theta_i-\langle \theta\rangle)^2}{\sum_i w_i}.
\end{equation}
These give the inferred values and uncertainties for $\Theta$.

\subsubsection{Practical implementation}
In the formalism described above, the likelihood function and inference framework are defined in a general form.
To apply this method in practice, we must specify three remaining ingredients:
the prior distributions for the stellar parameters, the grid of theoretical isochrones over which the likelihood is evaluated, and the initial mass function.
In this section, we describe how these choices are made for the MSTO sample, guided by stellar evolution theory, the observed color-magnitude diagrams, and existing knowledge of the Galactic halo.

The sensitivity of the MSTO to stellar age is a well-established result of stellar evolution theory.
Because the main sequence represents the longest-lived evolutionary phase,
the point at which stars leave the main sequence provides a direct clock for stellar populations.
As a consequence, the MSTO position in a color-magnitude diagram is expected to respond strongly to age variations.

\begin{figure*}[th!]
    \begin{center}
    \includegraphics[width=\textwidth, keepaspectratio]{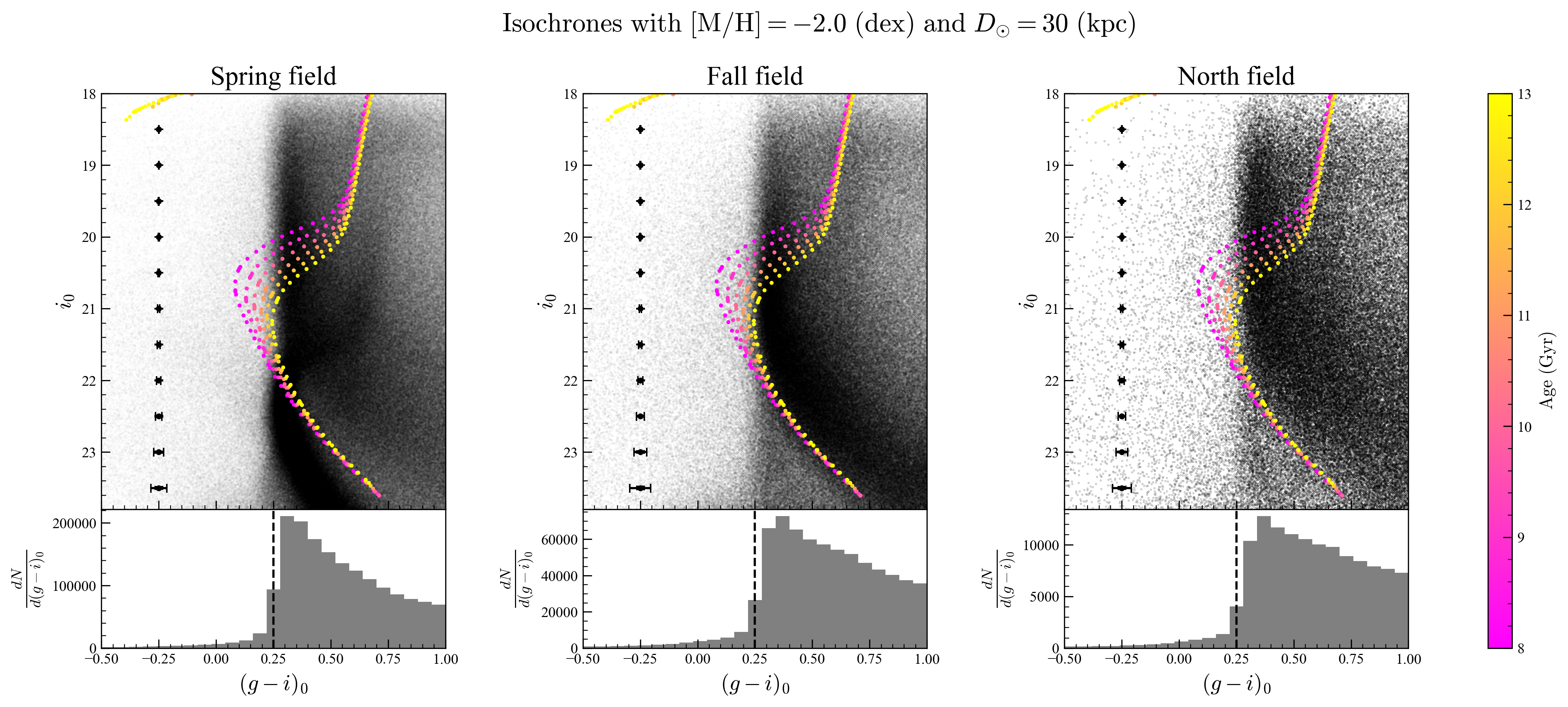}
    \end{center}
    \caption{
    Color-magnitude diagrams toward the Spring (left), Fall (middle), and North (right) fields in the HSC-SSP footprint.
    For each field, the top panel shows the dereddened $(g-i)_{0}$ versus $i_{0}$ diagram, where black points indicate the observed stars.
    The black dots with error bars show typical photometric uncertainties as a function of $i_{0}$.
    The bottom panel shows the $(g-i)_{0}$ color histogram, with a bin width equal to the color error at $i_{0}=24.5$.
    The vertical black dashed line marks the color at which the difference in counts between adjacent bins is maximal.
    In all three fields this produces a sharp edge at $(g-i)_{0}\simeq 0.25$.
    Colored dots show isochrone points at fixed metallicity $[\mathrm{M/H}]=-2.0$, ages from 8 to 13~Gyr in steps of 1~Gyr, and distance from the Sun $30~\mathrm{kpc}$, illustrating the age dependence of the MSTO position.
    {Alt text: A two-panel figure. The top panel shows color–magnitude diagrams ($(g-i)_{0}$ vs $i_{0}$) toward the Spring (left), Fall (middle), and North (right) fields. The bottom panel shows histograms of $(g-i)_{0}$ color for the same fields, with a sharp increase around $(g-i)_{0} \simeq 0.25$.}
    }
    \label{fig2}
\end{figure*}

We use the observed HSC color-magnitude diagrams to identify the location of the turnoff in the halo fields.
Figure~\ref{fig2} shows the color-magnitude diagrams for the Spring, Fall, and North fields, together with isochrone points for different stellar ages at a fixed metallicity of $[\mathrm{M/H}]=-2.0$.
The observed color distribution exhibits a sharp rise at $(g-i)_{0}\simeq 0.25$,
which is reproduced only by old isochrones with ages $\tau \gtrsim 10$~Gyr.
Younger models shift significantly blueward and fail to match this feature.
This behavior confirms that the MSTO population in the Milky Way stellar halo is dominated by old stars.
A similar conclusion was already reached by \citet{Unavane+1996}, who demonstrated that even a small intermediate-age population would produce a detectable excess of blue turnoff stars.

Based on the observed turnoff location, we define MSTO candidates using the color range
\begin{equation}
0.25 \le (g-i)_{0} \le 0.40,
\end{equation}
which encloses the turnoff feature in all three fields.
Within this region, the color-magnitude diagrams provides strong leverage on stellar age, motivating the use of an age prior restricted to old populations.

In contrast to its strong dependence on age, the position of the MSTO exhibits only a weak dependence on metallicity for the metal-poor halo stars since the contribution of metals to the stellar opacity is reduced at low metallicity. 
While variations in $[\mathrm{M/H}]$ do affect the detailed morphology of the turnoff, the broad-band CMD alone does not provide sufficient information to tightly constrain the metallicity distribution.
We therefore adopt an external prior on metallicity based on previous spectroscopic studies of halo stars.
In particular, we use the halo metallicity distribution function inferred by e.g., \citet{Ryan+1991, Schorck+2009}, which peaks near $[\mathrm{M/H}] \simeq -1.6$.

\begin{figure}[th!]
    \includegraphics[width=0.5\textwidth, keepaspectratio]{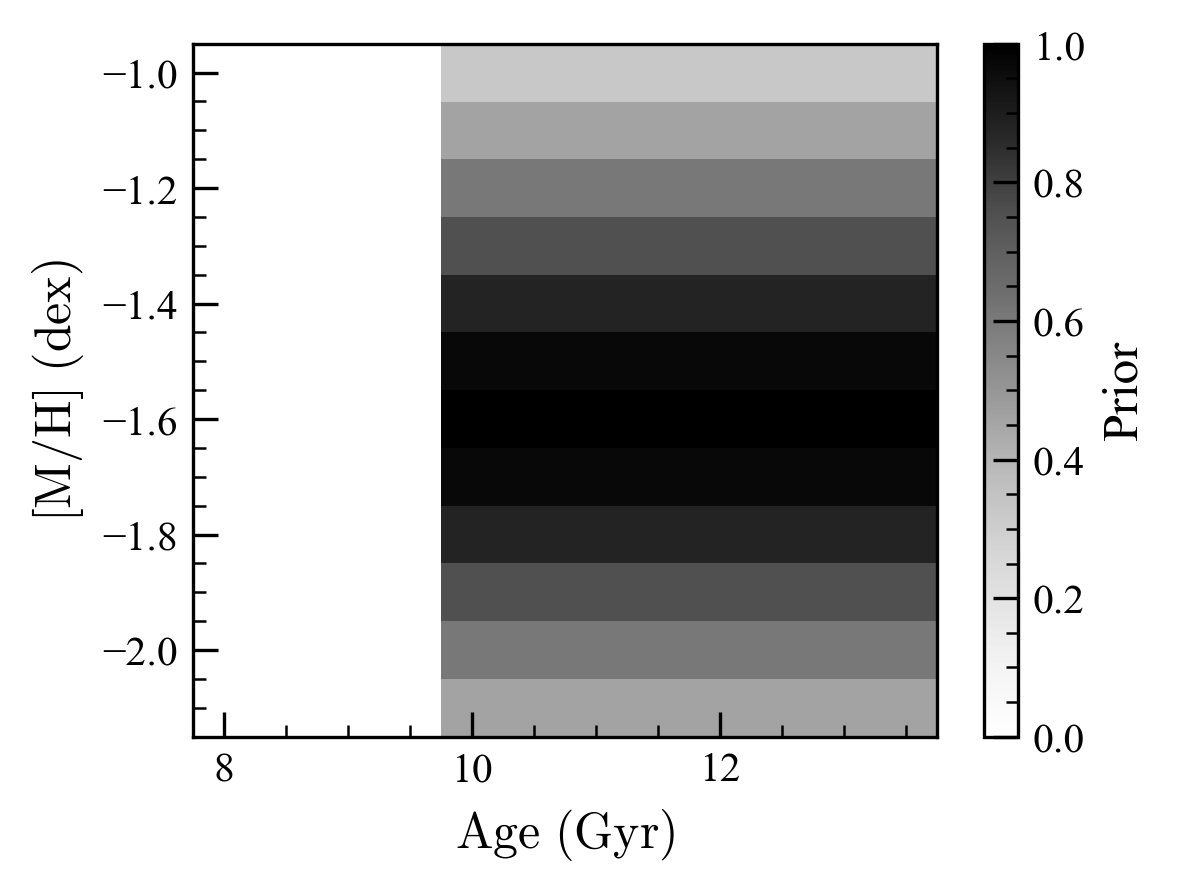}
    \caption{
    Adopted halo-motivated priors on stellar age and metallicity used for the
    isochrone-based distance estimation of the main-sequence turn-off sample.
    The age prior is uniform for old populations ($\tau \ge 10~\mathrm{Gyr}$)
    and smoothly suppressed toward younger ages by a Gaussian cutoff with
    $\sigma_{\tau}=0.1~\mathrm{Gyr}$.
    The metallicity prior is Gaussian, centered at $[\mathrm{M/H}]=-1.6$ with
    a dispersion of $0.4$~dex.
    Age and metallicity are assumed to be independent in the prior.
    {Alt text: A two-dimensional prior distribution in stellar age and metallicity. The age axis shows a distribution that is uniform for ages older than 10 Gyr and decreases toward younger ages with a Gaussian cutoff. The metallicity axis follows a Gaussian distribution centered at [M/H] = $-$1.6 with a dispersion of 0.4 dex.}
    }
    \label{fig3}
\end{figure}

We therefore adopt halo-motivated priors on age and metallicity, as illustrated in Figure~\ref{fig3}.
For the stellar age, we impose a prior that is uniform for old populations and rapidly suppressed toward younger ages,
\begin{equation}
p(\tau) \propto
\begin{cases}
\exp\!\left[-\dfrac{(\tau-\tau_{\rm cut})^2}{2\sigma_{\tau}^2}\right], & \tau < \tau_{\rm cut},\\[6pt]
1, & \tau \ge \tau_{\rm cut},
\end{cases}
\end{equation}
with $\tau_{\rm cut}=10~{\rm Gyr}$ and $\sigma_{\tau}=0.1~{\rm Gyr}$.
This form reflects the dominance of old stars in the Galactic halo while allowing for a smooth suppression toward younger ages rather than a hard cutoff.
For the metallicity, we adopt a Gaussian prior,
\begin{equation}
p([\mathrm{M/H}]) \propto 
\exp\!\left[
-\frac{([\mathrm{M/H}]-\mu_{\mathrm{M/H}})^2}{2\sigma_{\mathrm{M/H}}^2}
\right],
\end{equation}
with $\mu_{\mathrm{M/H}}=-1.6$ dex and $\sigma_{\mathrm{M/H}}=0.4$ dex.
We assume independence between age and metallicity in the prior, such that $p(\tau,[\mathrm{M/H}]) \propto p(\tau)\,p([\mathrm{M/H}])$.
These priors are adopted throughout the isochrone-based distance estimation.

Then, we evaluate the likelihood on a discrete grid in age, metallicity, and distance modulus.
We use the theoretical isochrone model from \citet{Bressan+2012}.
The age grid spans $\tau = 8.0$ to 13.5~Gyr in steps of 0.5~Gyr, yielding 12 grid points.
The metallicity grid spans $[\mathrm{M/H}] = -2.1$ to $-1.0$ dex, sampled linearly with a step size of 0.1~dex, also yielding 12 grid points.
The distance is parameterized in terms of the distance modulus $\mu$, which is sampled uniformly from $\mu = 10.0$ to 22.5 using 200 grid points.
This range corresponds to heliocentric distances from 1 to $\sim316$~kpc.
In evaluating the likelihood, we use only the portion of each isochrone from the main sequence to the sub-giant branch.
This ensures that the distance inference is based exclusively on the MSTO and adjacent evolutionary phases, which are the focus of this analysis.

For convenience, we adopt the Salpeter IMF, $\phi(M_\mathrm{ini}) \propto M_\mathrm{ini}^{-2.35}$ \citep{Salpeter+1955}.
For the old and metal-poor stellar populations considered here, MSTO stars correspond to a narrow range of initial masses, centered around $M_\mathrm{ini} \simeq 0.8\,M_{\odot}$.
Because our analysis is sensitive only to this limited mass range,
differences among commonly used IMFs, such as the Salpeter and Kroupa IMFs, are minimal in practice.
Consequently, the specific choice of IMF is not expected to have a significant impact on the results presented in this work.

\begin{figure}[th!]
    \includegraphics[width=0.5\textwidth, keepaspectratio]{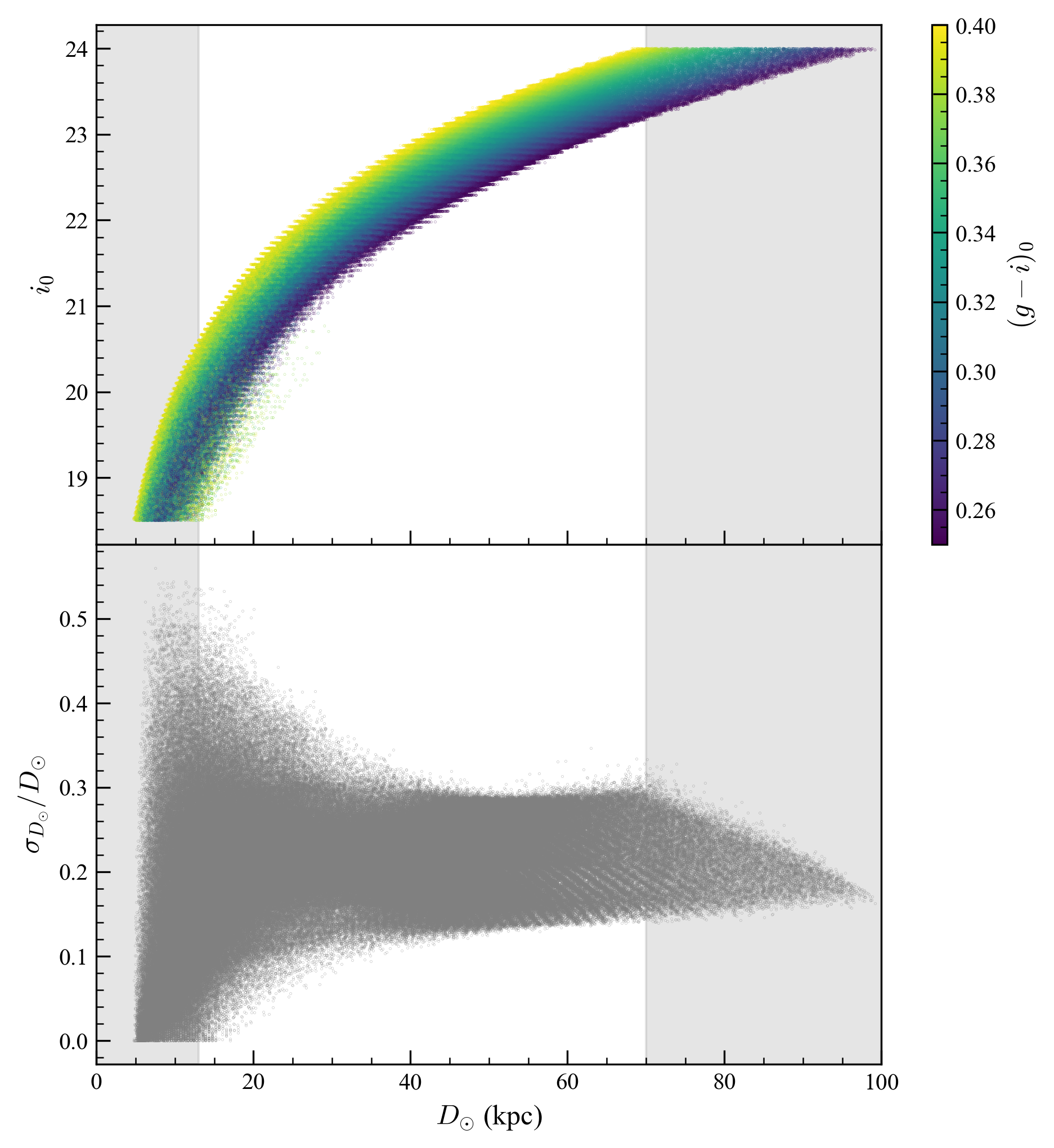}
    \caption{
    Posterior results for MSTO stars in the Spring field, shown as a function of heliocentric distance $D_{\odot}$.
    The top panel shows the observed de-reddened magnitude $i_{0}$.
    The color scale represents the $(g-i)_0$ color.
    The bottom panel shows the fractional distance uncertainty, $\sigma_{D_{\odot}}/D_{\odot}$, derived from the posterior distribution.
    The shaded gray area marks the magnitude-limited region from the $i_0$--$D_{\odot}$ relation in the top panel, which is excluded from our MSTO sample.
    The Fall and North fields show qualitatively similar behavior.
    {Alt text: A two-panel figure showing MSTO stars in the Spring field as a function of heliocentric distance $D_{\odot}$. The top panel shows the de-reddened magnitude $i_{0}$ versus $D_{\odot}$. The bottom panel shows the fractional distance uncertainty $\sigma_{D_{\odot}}/D_{\odot}$ versus $D_{\odot}$. 
    A gray-shaded region marks the magnitude-limited region defined by the $i_0$--$D_{\odot}$ relation. }
    }
    \label{fig4}
\end{figure}

\subsubsection{Sample selection}

We use the posterior distance estimates of MSTO stars to determine the range of heliocentric distance $D_{\odot}$ over which the sample can be regarded as effectively complete.
Figure~\ref{fig4} shows the posterior results for MSTO stars as a function of $D_{\odot}$. 
The bottom panel displays the fractional distance uncertainty, $\sigma_{D_{\odot}}/D_{\odot}$,
derived from the posterior distribution.
At small $D_\odot < 13~\mathrm{kpc}$, the fractional uncertainty increases, despite the small photometric errors.
In this regime, nearby main-sequence stars and more distant subgiant stars become degenerate in color-magnitude space, which broadens the posterior distribution in distance and increases the dispersion of the inferred mean distance.
Over an intermediate distance range, the fractional distance uncertainty remains
approximately constant at a level of $\sigma_{D_{\odot}}/D_{\odot} \simeq 0.2$, which corresponds to $\sigma_{\mu} \simeq 0.43$.
The top panel shows the observed de-reddened magnitude $i_{0}$ as a function of $D_{\odot}$.
For a given $D_\odot$, the MSTO sample must be complete in the $i_{0}$ direction in order to ensure an unbiased distance distribution.
The sharp truncation at faint magnitudes reflects the survey magnitude limit, which leads to an incompleteness of MSTO stars at large distances beyond 70 kpc.

Taking these effects together, we define the distance range
\begin{equation}
13 \leq D_{\odot}/{\rm kpc} \leq 70
\end{equation}
as the interval over which the MSTO sample can be treated as effectively complete.
In the following analysis, we therefore restrict the MSTO sample to this distance range when deriving the structural parameters of the MW's stellar halo.

\subsection{Stellar halo fitting}

The structure of the Galactic stellar halo is usually defined in Galactocentric coordinates through a three-dimensional number density profile $\nu(\boldsymbol{x}\mid\Theta)$, where $\Theta$ denotes a set of structural parameters such as radial slopes, breaks, or flattening, and $\boldsymbol{x}$ denotes the Galactocentric coodinates. 
In contrast, observational data are obtained in the heliocentric frame and consist of sky positions $(l,b)$ and distance moduli $\mu$, affected by measurement uncertainties and subject to incomplete sky coverage and selection effects.
Because of this mismatch, a direct transformation of the observed data into Galactocentric coordinates is generally invalid. 
Distance uncertainties smear the true radial distribution along the line of sight, while observational selection restricts the sampled volume in a highly non-uniform manner. 
If these effects are ignored, the inferred density profile is inevitably biased.

Therefore, the only consistent approach is to forward-model the Galactocentric stellar number density profile into the heliocentric observational space. 
In this framework, the stellar number density model $\nu(\boldsymbol{x}\mid\Theta)$ is mapped to the space of observables by explicitly accounting for coordinate transformations, distance modulus errors, and selection effects, and the model parameters are inferred by comparing the predicted distribution with the observed data through a likelihood function.
In the following, we first define the set of parameters, then describe the formulation. In the end, we describe the practical numerical implementation and parameter inference, validated using mock catalogs.

\subsubsection{Definition}

We begin by defining the halo model parameters $\Theta$ and the observed data $\mathcal{D}$.
The physical quantity of interest is the three--dimensional number density distribution of stellar halo stars defined in Galactocentric coordinates,
\begin{equation}
\nu(\boldsymbol{x}\mid\Theta),
\end{equation}
where $\boldsymbol{x}$ denotes the Galactocentric spatial position vector and $\Theta$ represents the set of halo structural parameters to be inferred.
The observed data for each star are given by
\begin{equation}
\mathcal{D}_{\mathrm{obs},i}
\equiv
\left\{
l_{i,\mathrm{obs}},
b_{i,\mathrm{obs}},
\mu_{i,\mathrm{obs}},
\sigma_{\mu,i}
\right\},
\end{equation}
where $(l_{i,\mathrm{obs}}, b_{i,\mathrm{obs}})$ are the Galactic longitude and latitude, $\mu_{i,\mathrm{obs}}$ is the observed distance modulus, and
$\sigma_{\mu,i}$ is its measurement uncertainty.
We introduce the corresponding latent true quantities
\begin{equation}
\mathcal{D}_{\mathrm{true},i}
\equiv
\left\{
l_{i,\mathrm{true}},
b_{i,\mathrm{true}},
\mu_{i,\mathrm{true}}
\right\}.
\end{equation}
The relationship between the observed and true quantities is specified by a measurement-error model, $E(\mathcal{D}_{\mathrm{obs},i}\mid\mathcal{D}_{\mathrm{true},i})$, which defines the conditional probability density of obtaining the observed data given the true values.
In this work, observational uncertainties in the angular coordinates are assumed to be negligible, while uncertainties in the distance modulus are explicitly modeled.
Specifically, the error model
$E(\mathcal{D}_{\mathrm{obs},i}\mid\mathcal{D}_{\mathrm{true},i})$
is given by
\begin{equation}
\delta(l_{i,\mathrm{obs}}-l_{i,\mathrm{true}})
\,
\delta(b_{i,\mathrm{obs}}-b_{i,\mathrm{true}})
\,
\mathcal{N}\!\left(
\mu_{i,\mathrm{obs}}
\mid
\mu_{i,\mathrm{true}},
\sigma_{\mu,i}^{2}
\right),
\end{equation}
where $\mathcal{N}(\mu_{i,\mathrm{obs}}\mid\mu_{i,\mathrm{true}},\sigma_{\mu,i}^{2})$ denotes a normal distribution with mean $\mu_{i,\mathrm{true}}$ and variance $\sigma_{\mu,i}^{2}$.

For a given set of halo parameters $\Theta$, the stellar number density model $\nu(\boldsymbol{x}\mid\Theta)$ specifies the spatial distribution of stars in Galactocentric coordinates, but the true heliocentric quantities $\mathcal{D}_{\mathrm{true}}$ are not directly observed.
We therefore define the likelihood for $\Theta$ by forward-modeling the stellar halo number density into the observational space and marginalizing over the latent true quantities,
\begin{equation}
\mathcal{L}(\Theta \mid \mathcal{D}_{\mathrm{obs}})
=
\int
p(\mathcal{D}_{\mathrm{true}} \mid \Theta)\,
E(\mathcal{D}_{\mathrm{obs}} \mid \mathcal{D}_{\mathrm{true}})
\, d\mathcal{D}_{\mathrm{true}},
\end{equation}
where $p(\mathcal{D}_{\mathrm{true}} \mid \Theta)$ denotes the distribution of the
true heliocentric quantities implied by the Galactocentric number density model
$\nu(\boldsymbol{x}\mid\Theta)$.

\subsubsection{Forward modeling framework}
We introduce the forward-modeling framework to consider actual observations.
For convenience, we define the line--of--sight intensity implied by the stellar halo number density model as
\begin{equation}
\lambda(l,b,\mu_{\mathrm{true}}\mid\Theta)
\equiv
\nu\!\left(\boldsymbol{x}(l,b,\mu_{\mathrm{true}})\mid\Theta\right)
\,
\frac{\ln 10}{5}\, d(\mu_{\mathrm{true}})^3 \cos b .
\end{equation}
Then the single--star likelihood can then be written as
\begin{equation}
\mathcal{L}_i(\Theta)
=
\frac{
\displaystyle
\int
\lambda(l_i,b_i,\mu_{\mathrm{true}}\mid\Theta)\,
\mathcal{N}\!\left(
\mu_{i,\mathrm{obs}}
\mid
\mu_{\mathrm{true}},
\sigma_{\mu,i}^{2}
\right)
\, d\mu_{\mathrm{true}}
}{
Z_{\mathrm{eff}}(\Theta)
}.
\end{equation}
The effective normalization ensures that the likelihood is properly normalized in the presence of incomplete survey coverage and is given by
\begin{equation}
Z_{\mathrm{eff}}(\Theta)
=
\int
\lambda(l,b,\mu_{\mathrm{true}}\mid\Theta)\,
P_{\mathrm{sel}}(l,b,\mu_{\mathrm{true}})
\, dl\, db\, d\mu_{\mathrm{true}} .
\end{equation}
Here
\begin{equation}
P_{\mathrm{sel}}(l,b,\mu_{\mathrm{true}})
=
\int
S(\mathcal{D}_{\mathrm{obs}})
\,
\mathcal{N}\!\left(
\mu_{\mathrm{obs}}
\mid
\mu_{\mathrm{true}},
\sigma_{\mu}^{2}
\right)
\, d\mu_{\mathrm{obs}}
\end{equation}
denotes the probability that a star with true distance modulus $\mu_{\mathrm{true}}$ enters the observed sample.

Assuming statistical independence between stars, the full log--likelihood is
\begin{equation}
\log\mathcal{L}(\Theta)
=
\sum_{i=1}^{N}\log\mathcal{L}_i(\Theta).
\end{equation}

\subsubsection{Practical Implementation}

In practice, the likelihood expressions derived above are evaluated numerically.
Because the HSC--SSP footprint is narrow, the angular dependence of the effective normalization is weak.
We therefore treat the $dl\,db$ part of the normalization as approximately constant and reduce the problem to one-dimensional integrals along the distance modulus direction.
For convenience, we introduce the line-of-sight intensity for the $i$-th star,
\begin{equation}
\lambda_i(\mu_{\mathrm{true}}\mid\Theta)
\equiv
\lambda(l_i,b_i,\mu_{\mathrm{true}}\mid\Theta),
\end{equation}
where $\lambda(l,b,\mu_{\mathrm{true}}\mid\Theta)$ was defined in the previous subsection.
For each star, we evaluate the numerator
\begin{equation}
N_i(\Theta)
=
\int
\lambda_i(\mu_{\mathrm{true}}\mid\Theta)\,
\mathcal{N}\!\left(
\mu_{i,\mathrm{obs}}
\mid
\mu_{\mathrm{true}},
\sigma_{\mu,i}^{2}
\right)
\, d\mu_{\mathrm{true}},
\end{equation}
and the corresponding normalization term
\begin{equation}
\Lambda_i(\Theta)
=
\int
\lambda_i(\mu_{\mathrm{true}}\mid\Theta)\,
P_{\mathrm{sel},i}(\mu_{\mathrm{true}})
\, d\mu_{\mathrm{true}},
\end{equation}
so that the per-star likelihood is given by
\begin{equation}
\mathcal{L}_i(\Theta)
=
\frac{N_i(\Theta)}{\Lambda_i(\Theta)}.
\end{equation}
The selection function is assumed to depend only on the observed distance modulus.
We adopt a top-hat window in $\mu_{\mathrm{obs}}$,
\begin{equation}
S(\mu_{\mathrm{obs}})=
\begin{cases}
1, &
\mu_{\mathrm{obs,min}}
\le
\mu_{\mathrm{obs}}
\le
\mu_{\mathrm{obs,max}},\\
0, & \text{otherwise}.
\end{cases}
\end{equation}
The corresponding inclusion probability is
\begin{equation}
P_{\mathrm{sel},i}(\mu_{\mathrm{true}})
=
\int
S(\mu_{\mathrm{obs}})\,
\mathcal{N}\!\left(
\mu_{\mathrm{obs}}
\mid
\mu_{\mathrm{true}},
\sigma_{\mu,i}^{2}
\right)
\, d\mu_{\mathrm{obs}} ,
\end{equation}
Both $N_i(\Theta)$ and $\Lambda_i(\Theta)$ are evaluated numerically using Gauss-Legendre quadrature over a finite integration interval $[\mu_{\mathrm{true,min}},\mu_{\mathrm{true,max}}]$.

\subsubsection{Halo models, priors, and MCMC sampling}

In this work, we adopt simple spherically symmetric models for the stellar halo number density profile in order to quantify the radial density slope under controlled assumptions.
We consider two functional forms: a single power-law model and a double power-law model, both expressed as functions of the Galactocentric radius $r$.

For the single power-law model, the halo density profile is written as
\begin{equation}
\nu(r\mid \alpha)
\propto
r^{-\alpha},
\end{equation}
where $\alpha$ is the logarithmic slope of the density profile.
We impose the the constraint $\alpha>0$, excluding unphysical rising density profiles.

For the double power-law model, the density profile is given by
\begin{equation}
\nu(r\mid \alpha_{\mathrm{in}}, \alpha_{\mathrm{out}}, r_{\mathrm{break}})
\propto
\begin{cases}
r^{-\alpha_{\mathrm{in}}}, & r \le r_{\mathrm{break}},\\
r_{\mathrm{break}}^{\,\alpha_{\mathrm{out}}-\alpha_{\mathrm{in}}}\,r^{-\alpha_{\mathrm{out}}},
& r > r_{\mathrm{break}},
\end{cases}
\end{equation}
where $\alpha_{\mathrm{in}}$ and $\alpha_{\mathrm{out}}$ denote the inner and outer slopes, and $r_{\mathrm{break}}$ is the break radius.
The prefactor ensures continuity of the density profile at $r=r_{\mathrm{break}}$.
We require $\alpha_{\mathrm{in}}>0$ and $\alpha_{\mathrm{out}}>0$, and we restrict
$r_{\mathrm{break}}$ to lie within the radial range probed by the observed stellar sample, so that the break is constrained by the data.

In order to compare the goodness of fit among different models,
we use the Bayesian Information Criterion (BIC), defined as
\begin{equation}
\mathrm{BIC} = -2 \ln \mathcal{L}_{\max} + k \ln n,
\end{equation}
where $\mathcal{L}_{\max}$ is the maximum likelihood of the model,
$k$ is the number of free parameters, and $n$ is the number of data points.

Posterior distributions of the model parameters are sampled using the
affine-invariant ensemble Markov Chain Monte Carlo sampler implemented in the
\texttt{emcee} package \citep{Foreman+2013}.
This sampler employs the stretch--move algorithm, in which the ensemble of walkers evolves collectively and successive samples are not strictly independent.
Convergence is thus assessed using the integrated autocorrelation time estimated for each parameter.
Following standard recommendations for ensemble samplers, we discard the initial burn--in phase and retain only samples obtained after at least $50$ times the maximum autocorrelation time. 
This criterion ensures that the final posterior samples are effectively independent and provides a conservative convergence threshold.

\subsubsection{Application to the mock data}

\begin{figure}[ht!]
    \includegraphics[width=0.5\textwidth, keepaspectratio]{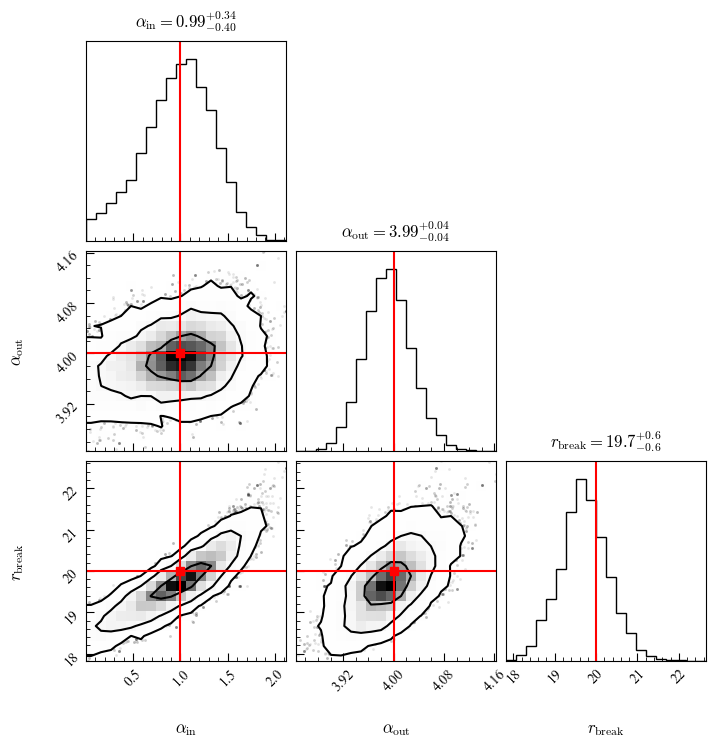}
    \caption{
    Posterior distributions of the double-power-law parameters
    $(\alpha_{\mathrm{in}}, \alpha_{\mathrm{out}}, r_{\mathrm{break}})$ inferred from the mock catalog.
    The red solid lines indicate the true input values $(\alpha_{\mathrm{in}}, \alpha_{\mathrm{out}}, r_{\mathrm{break}})
    =(1.0, 4.0, 20~\mathrm{kpc})$ used to generate the mock data.
    The contours represent the 1$\sigma$, 2$\sigma$, and 3$\sigma$ confidence regions.
    {Alt text: A corner plot showing posterior distributions of the parameters $(\alpha_{\mathrm{in}}, \alpha_{\mathrm{out}}, r_{\mathrm{break}})$. 
    The diagonal panels show one-dimensional distributions, while the off-diagonal panels show two-dimensional contours representing the 1$\sigma$, 2$\sigma$, and 3$\sigma$ confidence regions. Red solid lines indicate the true input values.}
    }
    \label{fig5}
\end{figure}

We apply the full forward-modeling framework to a mock stellar halo catalog in order to validate the inference method and to assess possible biases in the recovered halo parameters.
The mock catalog is generated from a spherically symmetric double power--law
density profile with known parameters, $(n_{\mathrm{in}},\, n_{\mathrm{out}},\, r_{\mathrm{break}}) = (1.0,\, 4.0,\, 20~\mathrm{kpc})$, motivated by previous models for the MW's stellar halo.
To closely mimic the observational conditions of the HSC-SSP survey, mock stars are first generated over the full sky according to the true halo model.
From this parent population, we select only those stars that fall within the North field in the HSC-SSP footprint, corresponding to an effective sky coverage of approximately $100~\mathrm{deg}^2$.
We further restrict the distance modulus range to $13 < D_{\odot}/\mathrm{kpc} < 70$ with $\sigma_{D_{\odot}} / D_{\odot} = 0.2$.
Using the relation between distance and distance modulus, this choice is equivalent to adopting a distance--modulus uncertainty $\sigma_{\mu}\sim0.43$.
The resulting mock catalog therefore reproduces both the spatial selection and the distance-error properties of the HSC-SSP data.

The mock sample is analyzed using the same likelihood function, prior assumptions, numerical integration scheme, and MCMC sampling strategy as applied to the observational data.
Posterior distributions of the parameters $(n_{\mathrm{in}}, n_{\mathrm{out}}, r_{\mathrm{break}})$ are shown in Figure~\ref{fig5}, where the red solid lines indicate the true input values.
All parameters are recovered within their $1\sigma$ credible intervals, demonstrating that the inference is unbiased within the statistical uncertainties.
The outer slope $n_{\mathrm{out}}$ and the break radius $r_{\mathrm{break}}$ are well constrained, while the posterior distribution of the inner slope $n_{\mathrm{in}}$ is noticeably broader.
This behavior is expected, as the restricted sky coverage and the adopted distance range limit the number of tracers probing the inner halo region.
Consequently, the present mock-data test highlights the reduced sensitivity to the inner density slope under HSC-SSP-like observational conditions.

We note that the mock catalog is intentionally simplified and does not include halo flattening, substructure, or spatially varying selection effects.
The purpose of this test is to validate the statistical framework and numerical implementation rather than to reproduce all complexities of the real stellar halo.
We expect that wider sky coverage or all--sky surveys would significantly improve constraints on the inner slope by increasing the number of tracers at small Galactocentric radii.
In such cases, however, the simplifying approximation adopted here such as treating the angular part of the likelihood normalization as approximately constant would break down.
A full treatment would require explicit modeling of the $(l,b)$ dependence of the line-of-sight integrals in both the likelihood numerator and the normalization for each star.

Overall, this application to mock data provides a consistency check of the entire analysis pipeline, from the forward-modeled likelihood and numerical integration to the MCMC sampling and convergence criteria, and establishes the range of validity of the present method.


\section{Results}

We first present the spatial distribution of MSTO stars in each survey field.
The distributions are shown in both the $(\mathrm{RA},\mathrm{Dec})$ and $(\mathrm{RA},D_\odot)$ planes to identify stellar substructures and to characterize the large-scale structure of the Galactic halo prior to modeling the smooth component.

\subsection{Spatial distribution of MSTO stars}

\begin{figure*}[t]
    \includegraphics[width=\textwidth]{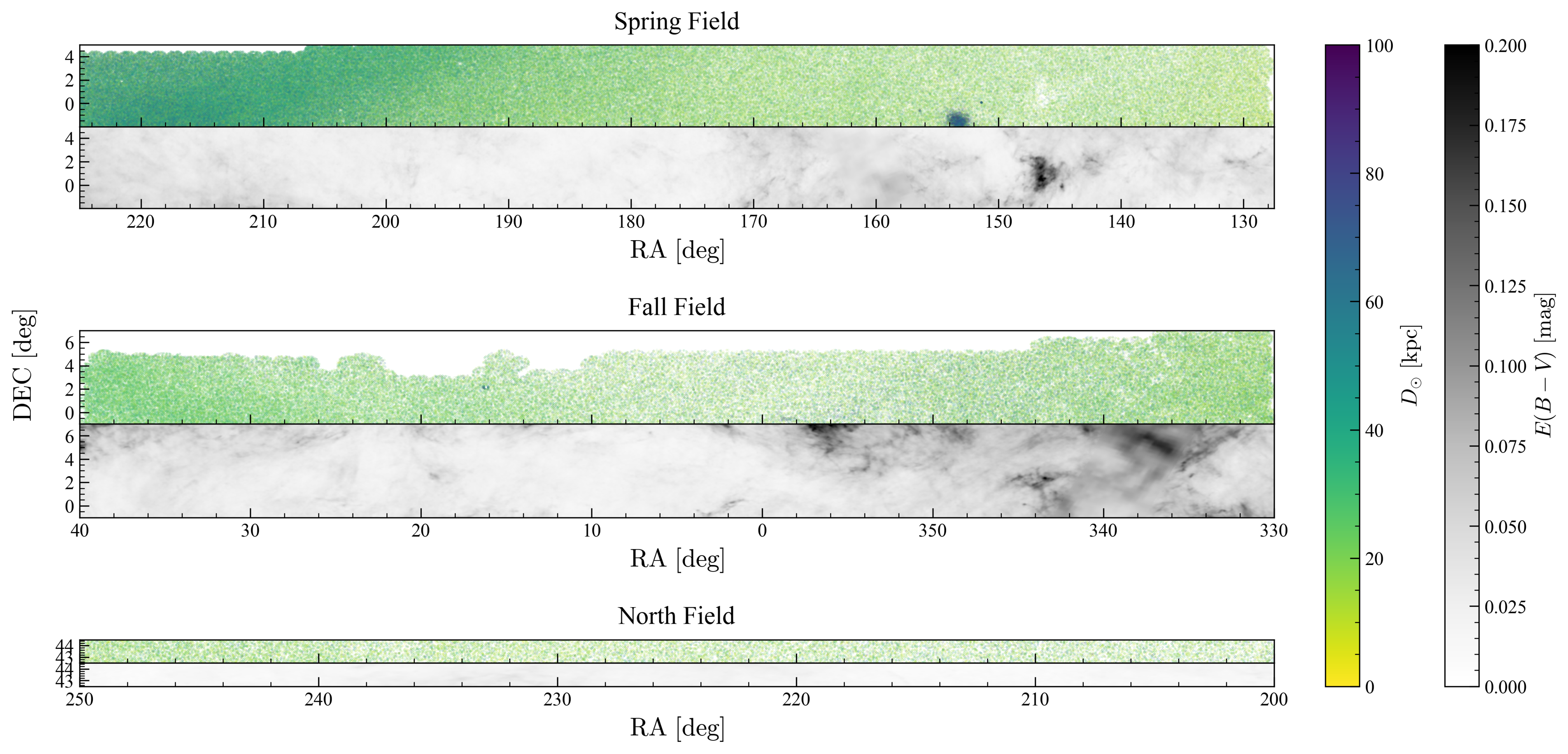}
    \caption{
    Spatial distribution of MSTO stars in the $(\mathrm{RA},\mathrm{Dec})$ plane for the Spring, Fall, and North fields.
    Each star is color-coded by heliocentric distance from 0 (yellow) to 100 kpc (purple) from the Sun.
    The corresponding extinction maps from \citet{Schlegel+1998} are shown below each panel.
    {Alt text: A multi-panel figure showing the spatial distribution of MSTO stars in equatorial coordinates (RA, Dec) for the Spring, Fall, and North fields. In each field, stars are plotted as a scatter distribution and are color-coded by heliocentric distance. Below each field, a corresponding extinction map is shown.}
    }
    \label{fig6}
\end{figure*}

Figure~\ref{fig6} presents the sky distribution of MSTO stars in the $(\mathrm{RA}, \mathrm{Dec})$ plane. 
Several known stellar systems and halo substructures are clearly identified in both the Spring and Fall fields. 
In the Spring field, a prominent overdensity associated with the Sagittarius stream is visible at $D_\odot \simeq$ $40$–$60\ \mathrm{kpc}$ over the range $\mathrm{RA} \simeq 200^\circ$–$225^\circ$. 
The Sextans dwarf spheroidal galaxy is also detected at $(\mathrm{RA},\mathrm{Dec}) = (162^\circ, -2^\circ)$. 
In addition, the globular cluster Palomar~3 (Sextans~C) appears clearly at $(\mathrm{RA},\mathrm{Dec}) = (151.375^\circ, 0.075^\circ)$, as does the recently discovered Sextans~II dwarf spheroidal galaxy at $(\mathrm{RA},\mathrm{Dec}) = (156.425^\circ, -0.600^\circ)$.
The Fall field likewise shows stars associated with the Sagittarius stream at heliocentric distances of $\sim 30$–$40\ \mathrm{kpc}$, together with a clear overdensity corresponding to the Hercules–Aquila Cloud \citep{Belokurov+2007} in the range $\mathrm{RA} \simeq 330^\circ$–$340^\circ$. 
In contrast, the North field exhibits no prominent overdensities in the projected stellar density, making it well suited for modeling the underlying smooth halo component.

\begin{figure*}[t]
    \includegraphics[width=\textwidth]{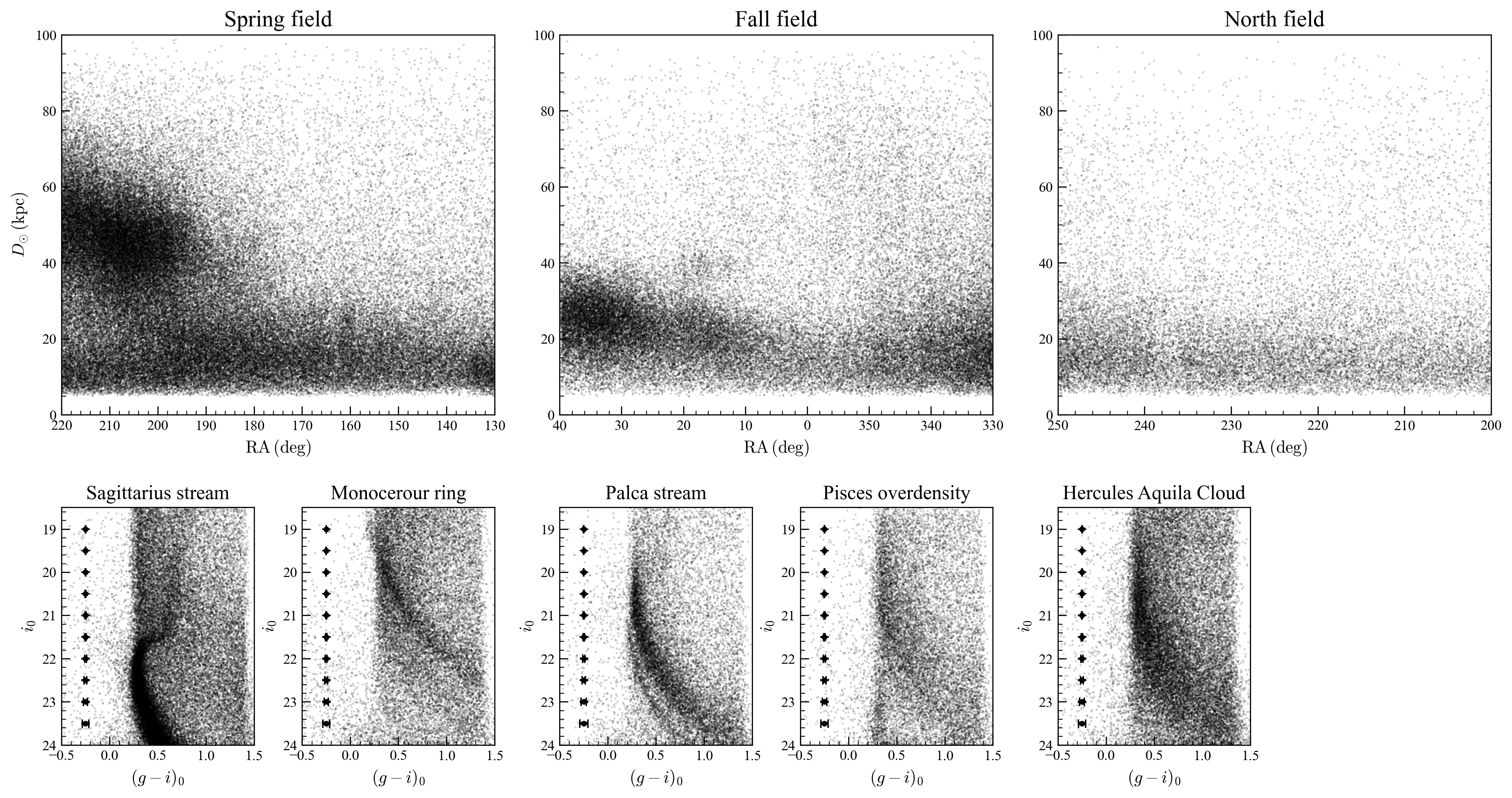}
    \caption{
    Distribution of MSTO stars in the $(\mathrm{RA}, D_\odot)$ plane for the Spring, Fall, and North fields.
    The top panels show the projected distributions in $(\mathrm{RA}, D_\odot)$, where the typical distance uncertainty of MSTO stars ($\sigma / D_\odot \sim 0.2$) significantly broadens the stellar distribution along the distance axis.
    The bottom panels present the color–magnitude diagrams toward the prominent substructures identified in the top panels. 
    The black dots with error bars show typical 1-sigma photometric uncertainties as a function of $i_{0}$.
    {Alt text: A multi-panel figure showing MSTO stars in the $(\mathrm{RA}, D_\odot)$ plane for the Spring, Fall, and North fields. In the top panels, stars are shown as a scatter distribution as a function of $(\mathrm{RA}, D_\odot)$ . The bottom panels show color–magnitude diagrams for selected substructures. Black points with error bars indicate typical photometric uncertainties as a function of $i_{0}$.}
    }
    \label{fig7}
\end{figure*}

Figure~\ref{fig7} shows the distribution of MSTO stars in the $(\mathrm{RA}, D_\odot)$ plane for the three survey fields: Spring ($130^\circ < \mathrm{RA} < 220^\circ$, $2^\circ < \mathrm{Dec} < 4^\circ$), Fall ($330^\circ < \mathrm{RA} < 400^\circ$, $-1^\circ < \mathrm{Dec} < 1^\circ$), and North ($200^\circ < \mathrm{RA} < 250^\circ$, $42.5^\circ < \mathrm{Dec} < 44.5^\circ$). Owing to the distance uncertainty of individual MSTO stars ($\sigma_{D_\odot } / D_\odot \sim 0.2$), the stellar distribution is significantly broadened along the distance axis. Nevertheless, several coherent overdensities are again clearly identifiable, particularly in the Spring and Fall fields.

In the Spring field, a prominent overdensity associated with the Sagittarius stream is identified at $\mathrm{RA} \simeq 190^\circ$–$220^\circ$ and $D_\odot \simeq 40$–$60\ \mathrm{kpc}$, with the corresponding CMD shown in the leftmost panel of the bottom row. At smaller distances, an overdensity at $\mathrm{RA} \simeq 130^\circ$–$135^\circ$ and $D_\odot \simeq 10$–$20\ \mathrm{kpc}$ is identified with the so-called Monoceros ring \citep{Newberg+2002}, with its CMD shown in the second panel from the left in the bottom row, while a more extended structure spanning $\mathrm{RA} \simeq 140^\circ$–$220^\circ$ over a similar distance range corresponds to the Virgo overdensity.

The Fall field also reveals multiple substructures in the $(\mathrm{RA}, D_\odot)$ plane. Stars associated with the Sagittarius stream are visible at $\mathrm{RA} \simeq 10^\circ$–$40^\circ$ and $D_\odot \simeq 20$–$30\ \mathrm{kpc}$. 
At larger distances, a diffuse overdensity at $\mathrm{RA} \simeq 330^\circ$–$360^\circ$ and $D_\odot \simeq 40$–$100\ \mathrm{kpc}$ is consistent with the Pisces overdensity, with the corresponding CMD shown in the fourth panel from the left in the bottom row. 
In addition, the Hercules–Aquila Cloud is detected at $\mathrm{RA} \simeq 330^\circ$–$350^\circ$ and $D_\odot \simeq 10$–$20\ \mathrm{kpc}$, with its CMD shown in the fifth panel from the left in the bottom row, and a narrower feature at $\mathrm{RA} \simeq 10^\circ$–$20^\circ$ and $D_\odot \sim 40\ \mathrm{kpc}$ is consistent with the Palca stream \citep{Shipp+2018}, with the corresponding CMD shown in the third panel from the left in the bottom row. 

In contrast, the North field exhibits a relatively smooth distribution in heliocentric distance, consistent with the absence of prominent substructures in both the projected density and the $(\mathrm{RA}, D_\odot)$ representations, supporting its use as a control field for modeling the smooth halo component.

\subsection{Density profile of the smooth stellar halo}

Based on the results presented in the previous sections, the North field provides the most suitable region for characterizing the smooth stellar halo of the MW and is treated as representative of its global, large–scale structure. 

However, as reported in Paper~I, there is evidence for a diffuse stellar overdensity, the so-called the Bo\"otes overdensity at a heliocentric distance of approximately $60\ \mathrm{kpc}$ in this direction. Although this feature does not appear as a prominent localized overdensity in the projected density map, its possible presence motivates a careful separation between the smooth halo component and more distant, low surface brightness structures. This consideration directly informs our analysis strategy. Thus, to disentangle the smooth halo component from potential distant substructures and to evaluate the robustness of the inferred halo parameters, we perform the analysis in two stages, using different heliocentric distance ranges.

Our modeling approach is further guided by the limited sky coverage of the North field, which spans only $\sim 100\ \mathrm{deg}^2$. Owing to this restricted spatial coverage, the data do not provide sufficient leverage to meaningfully constrain halo models with additional degrees of freedom, such as non–spherical geometries or multi–component profiles. We therefore adopt simple, spherically symmetric power–law models (single power-law and double powar-law model) as a first–order description of the global halo structure. 
To assess whether increased model complexity is warranted by the data, we compare a single power–law model and a broken power–law model using the BIC.

\begin{figure}[h!]
    \includegraphics[width=0.5\textwidth]{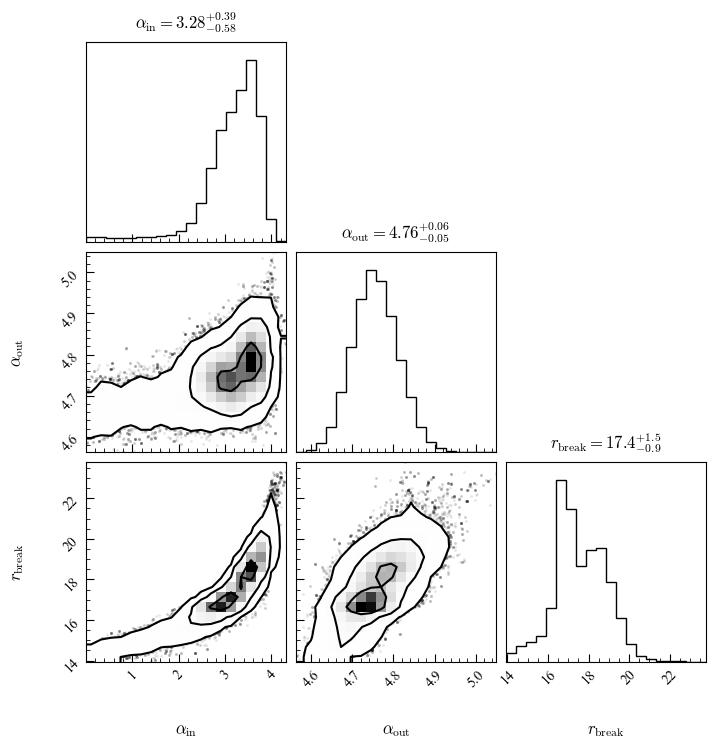}
    \caption{
    Posterior distributions of the smooth halo density parameters inferred from MSTO stars in the North field over the distance modulus range $15.6 \leq \mu \leq 18.5$, shown as a corner plot and assuming a spherically symmetric broken power--law model.
   The parameters are the inner slope $\alpha_{\mathrm{in}}$, the outer slope $\alpha_{\mathrm{out}}$, and the break radius $r_{\mathrm{break}}$.
    {Alt text: A corner plot showing posterior distributions of the parameters $(\alpha_{\mathrm{in}}, \alpha_{\mathrm{out}}, r_{\mathrm{break}})$ in the North field over the distance modulus range $15.6 \leq \mu \leq 18.5$. 
    The diagonal panels show one-dimensional distributions, while the off-diagonal panels show two-dimensional contours representing the 1$\sigma$, 2$\sigma$, and 3$\sigma$ confidence regions.}
   }
    \label{fig8}
\end{figure}

\begin{table*}[h!]
\caption{
Model comparison for the smooth halo density profile using MSTO stars within $15.6\leq \mu \leq 18.5$.
}
\centering
\label{tab:1}
\renewcommand{\arraystretch}{1.3}
\begin{tabular}{l p{4.2cm} p{6.5cm}@{\hspace{-8pt}} c}
\hline\hline
Model & Prior & Posterior & $\Delta \mathrm{BIC}$ \\
\hline
Double power law
&
\begin{tabular}[t]{@{}l@{}}
$0 < \alpha_{\mathrm{in}}$ \\
$0 < \alpha_{\mathrm{out}}$ \\
$0 \le r_{\mathrm{break}} \le 100\ \mathrm{kpc}$
\end{tabular}
&
$\alpha_{\mathrm{in}} = 3.28^{+0.39}_{-0.58}$ \par\smallskip
$\alpha_{\mathrm{out}} = 4.76^{+0.06}_{-0.05}$ \par\smallskip
$r_{\mathrm{break}} = 17.4^{+1.5}_{-0.9}\ \mathrm{kpc}$
&
$0.0$
\\
\hline
Single power law
&
$0 < \alpha$
&
$\alpha = 4.53^{+0.03}_{-0.03}$
&
$+39$
\\
\hline
\end{tabular}
\end{table*}

\begin{table*}[h!]
\caption{
Same as Table~\ref{tab:1}, but for MSTO stars within $16.0\leq \mu \leq 19.0$.
}
\centering
\label{tab:2}
\renewcommand{\arraystretch}{1.3}
\begin{tabular}{l p{4.2cm} p{6.5cm}@{\hspace{-8pt}} c}
\hline\hline
Model & Prior & Posterior & $\Delta \mathrm{BIC}$ \\
\hline
Double power law
&
\begin{tabular}[t]{@{}l@{}}
$0 < \alpha_{\mathrm{in}}$ \\
$0 < \alpha_{\mathrm{out}}$ \\
$0 \le r_{\mathrm{break}} \le 100\ \mathrm{kpc}$
\end{tabular}
&
$\alpha_{\mathrm{in}} = 4.69^{+0.04}_{-0.01}$ \par\smallskip
$\alpha_{\mathrm{out}} = 0.50^{+0.59}_{-0.35}$ \par\smallskip
$\alpha_{\mathrm{break}} = 54.8^{+1.3}_{-2.0}\ \mathrm{kpc}$
&
$0.0$
\\
\hline
Single power law
&
$0 < \alpha$
&
$\alpha = \mathrm{4.50}^{+0.03}_{-0.03}$
&
$+82$
\\
\hline
\end{tabular}
\end{table*}

We first conduct a fiducial analysis using MSTO stars within the distance modulus rage $15.6\leq \mu \leq 18.5$, which corresponds to the heliocentric distance range $13.1 < D_\odot < 50.1\ \mathrm{kpc}$. 
This range probes the inner–to–intermediate halo while minimizing contamination from distant, low surface brightness features suggested by previous studies. 
In order to run the mcmc, we set $\mu_{\mathrm{true,min}}=0$ and 
$\mu_{\mathrm{true,max}}=21.5$, which corresponds to the $D_{\odot,\mathrm{true,min}}=0.01~\mathrm{kpc}$ and 
$D_{\odot,\mathrm{true,min}}=200~\mathrm{kpc}$, respectively.
Within this distance range, we fit both the single power–law and broken power–law density models to the data.
The adopted priors and the resulting posterior constraints on the halo parameters are summarized in Table~\ref{tab:1}. 

As shown in Figure~\ref{fig8}, the outer density slope $n_{\mathrm{out}}$ is tightly constrained by the data, while the inner slope $n_{\mathrm{in}}$ is constrained with lower precision due to the limited coverage at small heliocentric distances, but remains data-informed rather than prior-dominated. 
Within this distance range, the BIC comparison favors the broken power-law model over the single power-law model.

\begin{figure}[h!]
    \includegraphics[width=0.5\textwidth]{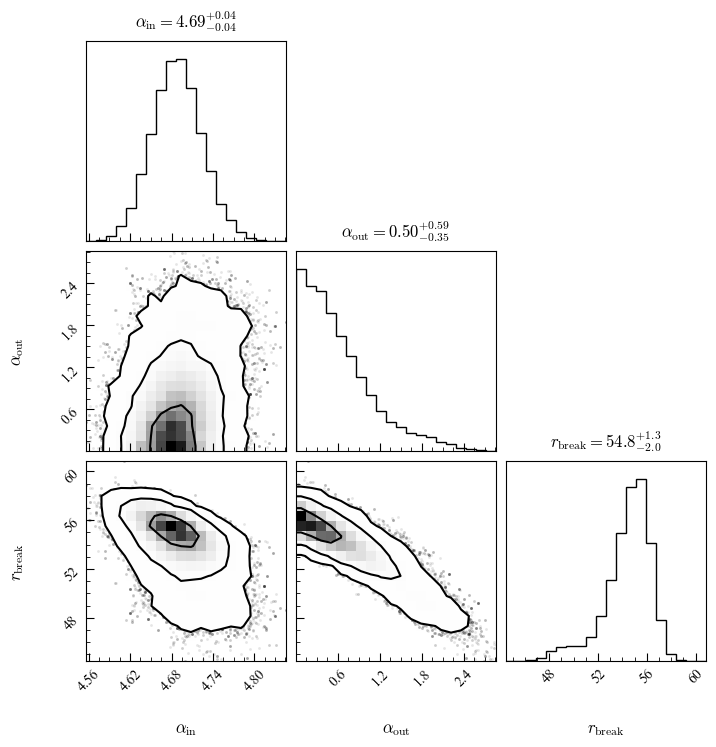}
    \caption{
    Posterior distributions of the smooth halo density parameters inferred from MSTO stars in the North field over the distance modulus range $16.0 \leq \mu \leq 19.0$, shown as a corner plot and assuming a spherically symmetric broken power-law model.
    The parameters are the inner slope $\alpha_{\mathrm{in}}$, the outer slope $\alpha_{\mathrm{out}}$, and the break radius $r_{\mathrm{break}}$.
    {Alt text: A corner plot showing posterior distributions of the parameters $(\alpha_{\mathrm{in}}, \alpha_{\mathrm{out}}, r_{\mathrm{break}})$ in the North field over the distance modulus range $16.0 \leq \mu \leq 19.0$. 
    The diagonal panels show one-dimensional distributions, while the off-diagonal panels show two-dimensional contours representing the 1$\sigma$, 2$\sigma$, and 3$\sigma$ confidence regions.}
    }
    \label{fig9}
\end{figure}

After establishing the preferred smooth halo model in the fiducial distance range, we extend the analysis to include MSTO stars out to $16.0\leq \mu \leq 19.0$, which corresponds to $15.8 < D_\odot < 63.1\ \mathrm{kpc}$.  
This extended analysis probes a more distant region of the halo and is designed to test whether the data favor the presence of additional halo components at large distances, such as a diffuse overdensity analogous to that suggested in previous work. In the extended analysis, we again compare the single and broken power-law models using the BIC to assess whether the increased distance coverage warrants additional model complexity. We summarize the adopted priors and the resulting posterior constraints on the halo parameters in Table~\ref{tab:2}. 

The obtained posterior distributions are shown in Figure~\ref{fig9}.
The inferred density slopes remain data-informed and are not dominated by the adopted priors.
In particular, the slope corresponding to the inner halo in the extended analysis is broadly consistent with the outer halo slope inferred from the previous fiducial sample ($13.1 < D_\odot < 50.1 \mathrm{kpc}$) .
In addition, the posterior distribution suggests the presence of a break in the density profile at $r_{\mathrm{break}} \simeq 54\ \mathrm{kpc}$. 
The BIC comparison favors the broken power--law model over the single power--law model, indicating that the data prefer a deviation from a simple, single--slope halo profile at large distances. 
This result suggests the presence of a diffuse halo substructure at $D_\odot \gtrsim 50\ \mathrm{kpc}$. This is independently consistent with Paper~I, which used main-sequence stars, although the distance in Paper~I was only roughly estimated by applying an isochrone filter.
Further investigation for this diffuse substructure in relation to the orbital motion of the Large Magellanic Cloud will be reported elsewhere as Suzuki et al. 2026b.


\section{Discussion}

\begin{figure*}[th!]
    \begin{center}
    \includegraphics[width=\textwidth, keepaspectratio]{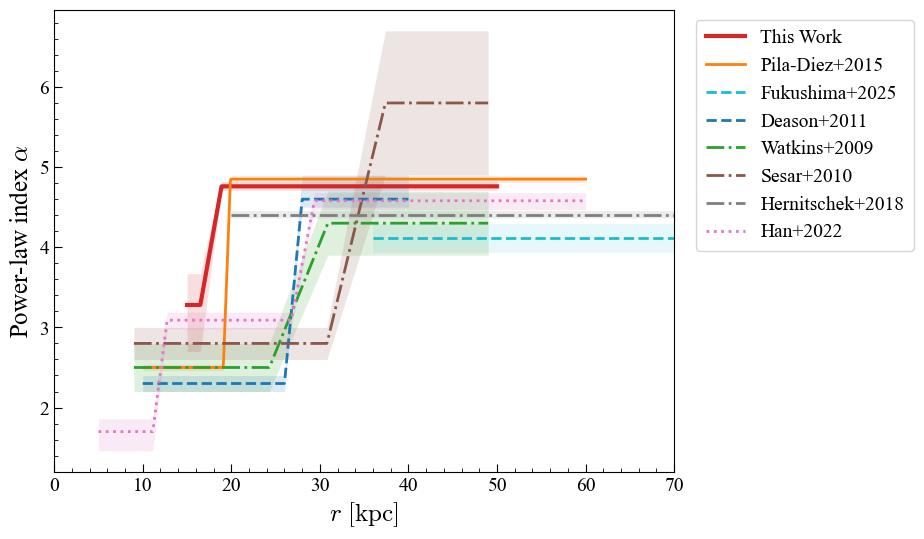}
    \end{center}
    \caption{
    Comparison of the inferred stellar halo density slopes with previous studies.
    Different line styles indicate different tracer populations.
    The solid line represents the density slope derived from MSTO stars, the dashed line from BHB stars, the dash--dotted line from RR Lyrae stars, and the dotted line from K-giants.
    The shaded regions in the corresponding colors indicate the 1$\sigma$ uncertainties for each work.
    {Alt text: A plot of stellar halo density slope $\alpha$ as a function of Galactocentric radius. Several lines are shown with different linestyles representing different tracer populations: solid for MSTO stars, dashed for BHB stars, dash-dotted for RR Lyrae stars, and dotted for K-giants. Shaded regions around each line indicate the 1$\sigma$ uncertainties.}
    }
    \label{fig10}
\end{figure*}

\subsection{Summary and comparison with previous works}
Using a Wide layer data of Subaru HSC-SSP, we have mapped the spatial distribution of abundant MSTO stars to trace the MW's stellar halo out to $D_\odot \sim70\ \mathrm{kpc}$. 
This wide-area mapping allows us to identify a number of previously known halo substructures such as Virgo overdensity \citep{Juric+2008} and Pisces overdensity \citep{Watkins+2009}, demonstrating that the MSTO population provides a powerful tracer of both the global halo and its embedded substructures.
To quantify the underlying smooth halo component, we focus on the North field, where the influence of prominent, previously identified substructures is minimized. 
Although this field represents only small footprints compared to other survey fields, it provides the cleanest available region for probing the smooth halo structure in our dataset. 
Also, as reported in Paper~I, the possible presence of a diffuse stellar overdensity at $D_\odot \sim 60\ \mathrm{kpc}$ is found in this direction. 
To avoid contamination from such distant, low–surface–brightness features, we therefore derive the smooth halo density profile using MSTO stars in the distance range $13.1 < D_\odot < 50.1\ \mathrm{kpc}$.
Within this range, we find that the stellar halo density is best described by a broken power-law profile with an inner slope of -3.28, an outer slope of -4.76, and a break radius of 17.4 kpc.

To place our result in context, we first compare the derived density slope with those reported in previous observational studies. A wide range of stellar-halo density slopes has been reported for different stellar populations, including BHB stars \citep{Deason+2011, Fukushima+2025}, RR Lyrae stars \citep{Watkins+2009, Sesar+2010, Hernitschek+2018}, MSTO stars \citep{PilaDiez+2015}, and K-giants \citep{Han+2022}, as summarized in Figure~\ref{fig10} (see also, Figure~14 in Li et al. 2026).
For the outer halo, measurements of the density slope show relatively good agreement across different studies, typically ranging between 4 and 5. In contrast, the inferred inner slope and break radius exhibit significant variation among previous works. One possible explanation for this diversity is the departure of the Milky Way stellar halo from spherical symmetry. Flattening or triaxiality can introduce line-of-sight–dependent density estimates, leading to apparent variations in the measured density slope.
Indeed, recent work by \citet{Han+2022} suggests that the stellar halo is tilted by approximately $25^\circ$ with respect to the Galactic plane toward the Sun and is well described by a triaxial ellipsoid with axis ratios of 10:8:7. In such a configuration, the derived density slope naturally depends on the line of sight, providing a physical explanation for part of the observed scatter among previous measurements.
Alternatively, the reported diversity may arise from systematic and statistical effects related to the choice of tracer population and analysis methodology. For example, samples of MSTO stars typically contain orders of magnitude more objects than samples of BHB or RR Lyrae stars, making them more susceptible to distance uncertainties, contamination, and selection effects if not properly modeled.
In the present work, the measured density slope is expected to be most robust because we explicitly incorporate the distance uncertainty of each individual star into the analysis. By mitigating both line-of-sight–dependent systematics and tracer-dependent statistical biases, our result provides a more reliable characterization of the smooth stellar halo density profile.

\subsection{Implications for the formation of the stellar halo}
Cosmological simulations and semi-analytic models within the $\Lambda$CDM paradigm establish a theoretical framework linking the stellar-halo density profile to a galaxy's accretion history. In general, the stellar halo density profile reflects the relative importance of early versus recent accretion events. Halos dominated by one or a few early, massive accretion events tend to develop steep density profiles, whereas systems assembled through multiple or prolonged accretion events exhibit shallower profiles.
This general behavior is illustrated by results from the Auriga simulations. Using MW–mass galaxies, \citet{Monachesi+2019} showed that stellar halos dominated by a single massive progenitor typically exhibit steep projected density profile $R^{-n}$ with slopes of order of $n=-3$, while systems assembled from multiple progenitors show shallower projected slopes, around $n=-2$.
Additional constraints are provided by the outer halo. Based on a semi-analytic model, \citet{Deason+2014} demonstrated that recent or ongoing satellite accretion preferentially populates the outer halo and results in a shallower outer density profile. In contrast, a steep outer-halo density slope indicates a lack of significant recent accretion.
In this context, our measured density profile provides direct insight into the MW's assembly history. The inner-halo density slope of $\alpha=-3.28$, corresponding to a projected slope of approximately $n=-2.28$, lies between the regimes associated with single massive mergers and prolonged accretion. At the same time, the very steep outer-halo slope of $\alpha \sim -4.7$ argues against substantial recent accretion. Taken together, these results suggest that the MW's stellar halo was assembled neither through a single dominant massive merger nor through extended recent accretion, but instead through multiple early accretion events.

The stellar halo number density slope should be interpreted as an integrated and non-unique diagnostic of a galaxy's assembly history. Substantial halo-to-halo scatter is present in both cosmological simulations and semi-analytic models, indicating that different combinations of progenitor mass, accretion time, internal structure, and orbital properties can lead to similar present-day density profiles \citep{Font+2011, Monachesi+2019}.
The formation of the stellar halo depends on multiple properties of the accreted satellites, including their masses, internal structures, and orbits, which collectively regulate how efficiently accreted material contributes to the present-day halo.
By construction, the present-day density distribution reflects the cumulative outcome of accretion, star formation, and subsequent dynamical evolution, effectively marginalizing over kinematics, chemical abundances, stellar ages, and time. As a result, the density slope alone cannot uniquely reconstruct individual merger events.
Additional factors such as the possible growth of an in-situ stellar component further contribute to the diversity of measured density slopes. 
The density slope therefore provides a global constraint on the overall accretion history of the Galaxy, but does not allow for a detailed reconstruction of specific mergers.

Looking ahead, wide-area deep photometric surveys will play a crucial role in constraining the global structure of the stellar halo. Ongoing surveys such as the Ultraviolet Near-infrared Optical Northern Survey (UNIONS) \citep{Gwyn+2025}, which covers the northern sky, and the Vera C. Rubin Observatory Legacy Survey of Space and Time (LSST) \citep{Ivezic+2019}, which will map the southern sky, will provide photometric data of comparable depth to HSC-SSP over much larger areas. 
These data will enable detailed studies of the halo’s overall shape, radial structure, and spatial substructures, offering new constraints on the MW's global assembly history.
In parallel, current multi-object spectroscopic facilities, including Subaru Prime Focus Spectrograph (PFS) \citep{Takada+2014} and the Dark Energy Spectroscopic Instrument (DESI) \citep{Schlafly+2023}, will provide extensive kinematic and chemical information for halo stars. 
Also, the astrometric satellite Gaia is expected to release DR4 around 2026 and DR5 around 2030, which will provide highly precise measurements of proper motions and parallaxes.
The combination of wide-field photometry, large spectroscopic samples, and precise astrometry will allow the stellar halo to be studied not only in terms of its density distribution, but also in velocity and chemical space, paving the way toward a more complete picture of the MW's formation.


\section{Conclusion}
In this paper, we have obtained the stellar halo density profile of the MW within 70 kpc from the Sun using the deep and wide imaging data taken from the Subaru/HSC.
We have constructed the catalog of MSTO stars as halo tracers based on theoretical isochrone models, and developed a forward-modeling framework that explicitly incorporates distance uncertainties to derive the structure parameter of the MW.
Using this method, we have robustly constrained the structural parameters of the MW's stellar halo and demonstrated that it is well described by a double power-law density profile, with an inner slope of $-3.28$, an outer slope of $-4.76$, and a break radius of 17.4~kpc. 
Comparison the inferred density slope with results from MW-like cosmological simulations, we conclude that the MW was build-up by a few early massive accretion events, one of them corresponds to Gaia Enceladus/Sausage.
Ongoing wide-field photometric surveys such as UNIONS and LSST as well as wide-field spectroscopic survey such as PFS and DESI will allow us to constrain the structure of the MW's stellar halo.


\begin{ack}
This work is supported in part by JST SPRING (No. JPMJSP2114) and JST, the establishment of university fellowships towards the creation of science technology innovation (No. JPMJFS2102), and the Graduate Program on Physics for the Universe (GP-PU), Tohoku University for YS, JSPS Grant-in-Aid for Scientific Research and MEXT Grant-in-Aid for Scientific Research (No. 18H04359, 21H05448, 24K00669 and 25H00394 for MC, 23K13098 and 25KJ0017 for SH). RFGW acknowledges support from Schmidt Sciences, through the generosity of Eric and Wendy Schmidt, by recommendation of the Schmidt Futures program. The HSC collaboration includes the astronomical communities of Japan and Taiwan and Princeton University.  The HSC instrumentation and software were developed by the National Astronomical Observatory of Japan (NAOJ), the Kavli Institute for the Physics and Mathematics of the Universe (Kavli IPMU), the University of Tokyo, the High Energy Accelerator Research Organization (KEK), the Academia Sinica Institute for Astronomy and Astrophysics in Taiwan (ASIAA), and Princeton University.  Funding was contributed by the FIRST program from the Japanese Cabinet Office, the MEXT, JSPS,  JST,  the Toray Science  Foundation, NAOJ, Kavli IPMU, KEK, ASIAA, and Princeton University. This paper makes use of software developed for the Large Synoptic Survey Telescope. We thank the LSST Project for making their code freely available. The Pan-STARRS1 (PS1) Surveys have been made possible through contributions of the Institute for Astronomy, the University of Hawaii, the Pan-STARRS Project Office, the Max-Planck Society and its participating institutes, the Max Planck Institute for Astronomy and the Max Planck Institute for Extraterrestrial Physics, The Johns Hopkins University, Durham University, the University of Edinburgh, Queen's University Belfast, the Harvard-Smithsonian Center for Astrophysics, the Las Cumbres Observatory Global Telescope Network Incorporated, the National Central University of Taiwan, the Space Telescope Science Institute, the National Aeronautics and Space Administration under Grant No. NNX08AR22G issued through the Planetary Science Division of the NASA Science Mission Directorate, the National Science Foundation under Grant No.AST-1238877, the University of Maryland, and Eotvos Lorand University (ELTE).
\end{ack}



\begin{thebibliography}{99}
\bibitem[Abbott et al.(2018)]{Abbott+2018} Abbott, T. M. C., Abdalla, F. B., Allam, S.\ 2018, ApJS, 239, 18
\bibitem[Bell et al.(2008)]{Bell+2008} Bell, E. F., Zucker, D. B. and Belokurov, V.\ 2008, ApJ, 680, 295
\bibitem[Belokurov et al.(2006)]{Belokurov+2006} Belokurov, V., Zucker, D. B., Evans, N. W.\ 2006, ApJ, 642, L137
\bibitem[Belokurov et al.(2007)]{Belokurov+2007} Belokurov, V., Evans, N. W., Bell, E. F. \ 2007, ApJ, 657, 89
\bibitem[Belokurov et al.(2018)]{Belokurov+2018} Belokurov, V., Erkal, D., Evans, N. W.\ 2018, MNRAS, 478, 611
\bibitem[Bosch et al.(2019)]{Bosch+2019} Bosch, J., Armstrong, R., Bickerton, S.\ 2019, PASJ, 71, 114
\bibitem[Bressan et al.(2012)]{Bressan+2012} Bressan, A., Marigo, P., Girardi, L\ 2012 MNRAS, 427, 127
\bibitem[Bullock \& Johnston(2005)]{Bullock+2005} Bullock, J. S., \& Johnston, K. V.\ 2005, ApJ, 635, 931
\bibitem[Carlin et al.(2012)]{Carlin+2012} Carlin, J. L., Majewski, S. R., Casetti-Dinescu, D. I.\ 2012, ApJ, 744, 25
\bibitem[Carollo et al.(2007)]{Carollo+2007} Carollo, D., Beers, T. C., Lee, Y. S.\ 2007, Nat, 450, 1020
\bibitem[Chambers et al.(2016)]{Chambers+2016} Chambers, K. C., Magnier, E. A., Metcalfe, N.\ 2016, arXiv:1612.05560
\bibitem[Cooper et al.(2010)]{Cooper+2010} Cooper, A. P., Cole, S., Frenk, C. S.\ 2010, MNRAS, 406, 744
\bibitem[Deason et al.(2011)]{Deason+2011} Deason, A. J., Belokurov, V., Evans, N. W. \ 2011, MNRAS, 416, 2903
\bibitem[Deason et al.(2014)]{Deason+2014} Deason, A. J., Belokurov, V., Koposov, S. E. \ 2014, ApJ, 787, 30
\bibitem[Eggen, Lynden-Bell and Sandage(1962)]{Eggen+1962} Eggen, O. J., Lynden-Bell, D., \& Sandage, A. R.\ 1962, ApJ, 136, 748
\bibitem[Font et al.(2011)]{Font+2011} Font, A. S., McCarthy, I. G., Crain, R. A. \ 2011, MNRAS, 416, 2802 
\bibitem[Foreman-Mackey et al.(2013)]{Foreman+2013} Foreman-Mackey, D., Hogg, D. W.,  Lang, D.\ 2013, PASP, 125, 306
\bibitem[Fukushima et al.(2025)]{Fukushima+2025} Fukushima, T., Chiba, M., and Tanaka, M. \ 2025, PASJ, 77, 178
\bibitem[Gywn et al.(2025)]{Gwyn+2025} Gwyn, S., McConnachie, A., W., Cuillandre, J. \ 2025, AJ, 170, 324
\bibitem[Han et al.(2022)]{Han+2022} Han, J. J., Conroy, C., Johnson, B. D. \ 2022, AJ, 164, 249
\bibitem[Helmi et al.(2018)]{Helmi+2018} Helmi, A., Babusiaux, C., Koppelman, H. H.\ 2018, Nature, 563, 85
\bibitem[Hernitschek et al.(2018)]{Hernitschek+2018} Hernitschek, N., Cohen, J. G., Rix, H. \ 2018, ApJ, 859, 31 
\bibitem[Ivezi\'c et al.(2008)]{Ivezic+2008} Ivezi\'c, \v{Z}., Sesar, B., Juri\'c, M.\ 2008, ApJ, 684, 287
\bibitem[Ivezi\'c et al.(2019)]{Ivezic+2019} Ivezi{\'c}, \v{Z}., Kahn, S. M., Tyson, J. A. \ 2019, ApJ, 873, 111 
\bibitem[Juri\'c et al.(2008)]{Juric+2008} Juri\'c, M., Ivezi\'c, \v{Z}., Brooks, A.\ 2008, ApJ, 673, 864
\bibitem[Juri\'c et al.(2017)]{Juric+2017} Juri\'c, M., Kantor, J., Lim, K. T.\ 2017, ASPC, 512, 279
\bibitem[Koppelman et al.(2019)]{Koppelman+2019} Koppelman, H. H., Helmi, A., Massari, D.\ 2019, A\&A, 631, L9
\bibitem[Kordopatis et al.(2023)]{Kordopatis+2023} Kordopatis, G., Ibata, R., Famaey, B.\ 2023, A\&A, 669, A104
\bibitem[Li et al.(2026)]{Li+2026} Li, S., Wang, W., Koposov, S. E., \ 2026, ApJ, 999, 108
\bibitem[Magnier et al.(2013)]{Magnier+2013} Magnier, E. A., Schlafly, E. F., Finkbeiner, D. P.\ 2013, ApJS, 205, 20
\bibitem[Monachesi et al.(2019)]{Monachesi+2019} Monachesi, A., G{\'o}mez, F. A., Grand, R. J. J. \ 2019, MNRAS, 485, 2589 
\bibitem[Myeong et al.(2019)]{Myeong+2019} Myeong, G. C., Vasiliev, E., Iorio, G., \ 2019, 488, 1235
\bibitem[Newberg et al.(2002)]{Newberg+2002} Newberg, H. J., Yanny, B., Rockosi, C. \ 2002, ApJ, 569, 245
\bibitem[Newberg et al.(2006)]{Newberg+2006} Newberg, J. H., \& Yanny, B. 2015, JPhCS, 47, 195
\bibitem[Nie et al.(2015)]{Nie+2015} Nie, J. D., Smith, M. C., Belokurov, V.\ 2015, ApJ, 810, 153
\bibitem[Pila-Di\'ez et al.(2015)]{PilaDiez+2015} Pila-Di\'ez, B., de Jong, J. T. A., Kuijken, K., van der Burg, R. F. J., \& Hoekstra, H. 2015, A\&A, 579, A38
\bibitem[Ryan et al.(1991)]{Ryan+1991} Ryan, S. G.\& Norris, J. E.\ 1991, AJ, 101, 1865
\bibitem[Salpeter(1955)]{Salpeter+1955} Salpeter, E. E., 1955, ApJ, 121, 161S
\bibitem[Schlafly et al.(2012)]{Schlafly+2012} Schlafly, E. F., Finkbeiner, D. P., Juri\'c, M.\ 2012, ApJ, 756, 158
\bibitem[Schlafly et al.(2023)]{Schlafly+2023} Schlafly, E. F., Kirkby, D., Schlegel, D. J. \ 2023, AJ, 166, 259
\bibitem[Schlegel et al.(1998)]{Schlegel+1998} Schlegel, D. J., Finkbeiner, D. P., \& Davis, M. 1998, ApJ, 500, 525
\bibitem[Searle \& Zinn(1978)]{Searle+1978} Searle, L., \& Zinn, R. 1978, ApJ, 225, 357
\bibitem[Sesar et al.(2007)]{Sesar+2007} Sesar, B., Ivezi\'c, \v{Z}., Lupton, R. H.\ 2007, AJ, 134, 2236
\bibitem[Sesar et al.(2010)]{Sesar+2010} Sesar, B., Ivezi{\'c}, \v{Z}, Grammer, S. H. \ 2010, ApJ, 708, 717  
\bibitem[Sesar et al.(2017)]{Sesar+2017} Sesar, B., Hernitschek, N., Dierickx, M. I. P., Fardal, M. A., \& Rix, H.-W. 2017, ApJ, 844, L4
\bibitem[Sch{\"o}rck et al.(2009)]{Schorck+2009} Sch{\"o}rck, T., Christlieb, N., Cohen, J. G. \ 2009, A\&A, 507, 817
\bibitem[Shipp et al.(2018)]{Shipp+2018} Shipp, N., Drlica-Wagner, A., Balbinot, E. \ 2018, ApJ, 862, 114
\bibitem[Springel et al.(2005)]{Springel+2005} Springel, V., White, S. D. M., Jenkins, A.\ 2005, Nature, 435, 629
\bibitem[Suzuki et al.(2024)]{Suzuki+2024} Suzuki, Y., Chiba, M., Komiyama, Y.\ 2024, PASJ, 76, 205
\bibitem[Takada et al.(2014)]{Takada+2014} Takada, M., Ellis, R. S., Chiba, M. \ 2014, PASJ, 66, 1
\bibitem[Tonry et al.(2012)]{Tonry+2012} Tonry, J. L., Stubbs, C. W., Lykke, K. R.\ 2012, ApJ, 750, 99
\bibitem[Unavane et al.(1996)]{Unavane+1996} Unavane, M., Wyse, R. F. G., Gilmore, G. \ 1996, 278, 727
\bibitem[Vasiliev et al.(2021)]{Vasiliev+2021} Vasiliev, E., Belokurov, V., \& Erkal, D. 2021, MNRAS, 501, 2279
\bibitem[Vivas et al.(2001)]{Vivas+2001} Vivas, A. K., Zinn, R., Andrews, P.\ 2001, ApJ, 554, L33
\bibitem[Watkins et al.(2009)]{Watkins+2009} Watkins, L. L., Evans, N. W., Belokurov, V.\ 2009, MNRAS, 398, 1757
\bibitem[White \& Rees(1978)]{White+1978} White, S. D. M., \& Rees, M. J. 1978, MNRAS, 183, 341
\bibitem[Xue et al.(2015)]{Xue+2015} Xue, X. , Rix, H., Ma, Z. \ 2015, 809, 144
\bibitem[Yanny et al.(2009)]{Yanny+2009} Yanny, B., Rockosi, C., Newberg, H., J.\ 2009, AJ, 137, 4377
\bibitem[York et al.(2000)]{York+2000} York, D. G., Adelman, J., Anderson, J. E.\ 2000, AJ, 120, 1579
\bibitem[Zhao et al.(2012)]{Zhao+2012} Zhao, G., Zhao, Y., Chu, Y.\ 2012, RAA, 12, 723Z
\bibitem[Zwitter et al.(2010)]{Zwitter+2010} Zwitter, T., Matijevi{\v{c}}, G., Breddels, M. A.\ 2010, A\&A, 522, 54
\end{thebibliography}
\end{document}